\documentclass[useAMS,usenatbib]{mn2e}
\usepackage[dvips]{graphicx}
\usepackage{longtable}

\title[Optical and IR observations of SN~2002dj ]{Optical and IR observations of SN~2002dj: some possible common properties of fast expanding SNe~Ia }
\author[G. Pignata]{G. Pignata,$^{1,2}$\thanks{E-mail:
gpignata@das.uchile.cl} S. Benetti,$^3$ P. A. Mazzali,$^{4,5}$ R. Kotak,$^{6}$ F. Patat,$^7$ P. Meikle,$^{8}$
\newauthor M. Stehle,$^4$  B. Leibundgut,$^{7}$ N. B. Suntzeff,$^{9}$ L. M. Buson,$^3$ E. Cappellaro,$^3$  
\newauthor A. Clocchiatti,$^{2}$ M. Hamuy,$^{1}$ J. Maza,$^{1}$ J. Mendez,$^{10}$ P. Ruiz-Lapuente,$^{10}$   
\newauthor M. Salvo,$^{11}$ B. P. Schmidt,$^{11}$ M. Turatto$^{3}$ and W. Hillebrandt$^{4}$\\
$^1$ Departamento de Astronom\'ia, Universidad de Chile, Casilla 36-D, Santiago, Chile\\
$^2$ Departamento de Astronom\'ia y Astrof\'isica, Pontificia Universidad Cat\'olica de Chile, Casilla 306, Santiago 22, Chile\\
$^3$ INAF Osservatorio Astronomico di Padova, Vicolo dell Osservatorio 5, I-35122 Padova, Italy\\
$^4$ Max-Planck-Institut  f\"ur Astrophysik, Karl-Schwarzschild-Str. 1, D-85741 Garching bei M\"unchen, Germany\\
$^5$ Osservatorio Astronomico di Trieste, Via Tiepolo 11, I-34131 Trieste, Italy\\
$^6$ Astrophysics Research Centre, School of Mathematics and Physics, Queen's University Belfast, Belfast BT7 1NN, United Kingdom\\
$^7$ European Southern Observatory,  Karl-Schwarzschild-Str. 2, D-85748 Garching bei M\"unchen,  Germany\\
$^8$ Blackett Laboratory, Imperial College London, Prince Consort Road, London SW7 2BW, United Kingdom\\
$^9$ Department of Physics, Texas A\&M University, College Station, TX 77843-4242\\
$^{10}$ Department of Astronomy, University of Barcelona, Marti i Franques 1, E-08028 Barcelona, Spain\\
$^{11}$ Research School of Astronomy and Astrophysics, Australian National University, Cotter Road, Weston Creek, ACT 2611, Australia\\
}

\begin{document}

\date{Accepted ...... Received .......; in original form .......}

\pagerange{\pageref{firstpage}--\pageref{lastpage}} \pubyear{2008}

\maketitle

\label{firstpage}

\begin{abstract}
As part of the European Supernova Collaboration we obtained extensive
photometry and spectroscopy of the type Ia SN~2002dj covering epochs
from 11 days before to nearly two years after maximum. Detailed optical
and near-infrared observations show that this object belongs to the
class of the high-velocity gradient events as indicated by Si, S and Ca
lines. The light curve shape and velocity evolution of SN~2002dj appear
to be nearly identical to SN~2002bo. The only significant difference is
observed in the optical to near-IR colours and a reduced spectral
emission beyond 6500~\AA.  For high-velocity gradient Type Ia
supernovae, we tentatively identify a faster rise to maximum, a more
pronounced inflection in the V and R light curves after maximum and a
brighter, slower declining late-time B light curve as common photometric
properties of this class of objects. They also seem to be characterized
by a different colour and colour evolution with respect to ``normal''
SNe~Ia. The usual light curve shape parameters do not distinguish these
events. Stronger, more blueshifted absorption features of intermediate-mass elements and
lower temperatures are the most prominent spectroscopic features of Type
Ia supernovae displaying high velocity gradients. It appears that these
events burn more intermediate-mass elements in the outer layers. Possible connections
to the metallicity of the progenitor star are explored.
\end{abstract}

\begin{keywords}
supernovae: general - supernovae: individual: SN~2002dj 
\end{keywords}

\section{Introduction} 
In the last decade type Ia Supernovae (SNe~Ia) have been extensively
used for cosmology yielding evidence that we live in an accelerating
Universe (Riess et al. 1998; Perlmutter et al. 1999).  Ongoing surveys
such as ESSENCE \citep{Miknaitis07,Wood-Vasey07} and SNLS \citep{Astier06} use the
relation between the shape of the light curve and its peak luminosity
\citep{Phillips93,Phillips99,Hamuy96,Riess96,Goldhaber01,Guy05,Prieto06}
to constrain the equation-of-state parameter for dark energy.  An
important caveat is that those relations  assume that SNe~Ia are a one
parameter family.

However, in recent years there has been growing evidence for the
observational diversity among SN~Ia and this is of prime interest in
their application as distance indicators. Subtle but unequivocal
differences between events with large wavelength coverage and dense
temporal sampling are evident in light curve shapes, colour evolutions,
luminosities, evolution of spectral lines and expansion velocities
derived from the line shifts.  The search for accurate correlations
between photometric and spectroscopic properties could improve the
luminosity calibration and help to shed light on the explosion mechanisms
and progenitor system. Recently, \citet{Benetti05} have identified
three classes of SNe~Ia based on their spectroscopic features. A similar
classification is also reported in \citet{Branch06}. Interestingly, two
of these classes are nearly indistinguishable in some of their
photometric parameters (e.g. $\Delta m_{15}$, Phillips et al. 1993); yet
they show clear spectroscopic differences.  High velocity gradient (HVG)
SNe are characterized by a fast decrease in their expansion velocity
over time as measured from the minimum of the Si~II (6355~\AA)
absorption line. On the other hand in the low velocity gradient (LVG)
SNe group, which represents the majority of SN~Ia, the evolution of the
Si~II (6355~\AA) velocity is smooth, the ejecta expansion is slower than
in HVG SNe. These spectroscopic differences may or may not affect the
relations used to calibrate the luminosities of SNe~Ia. Subtle
differences in colour may exist among the different classes and lead to
significant bias in reddening estimates.  The existence of spectroscopic
families suggests possible differences in the progenitor channels and/or
explosion mechanisms. \citet{Branch93} for example reported that SNe
characterized by high expansion velocities, tend to explode in late type
galaxies, pointing to progenitors arising from a young population.

In this paper we present the observations of the HVG SN~2002dj carried
out by the European Supernova Collaboration (ESC). SN~2002dj ($\alpha =
13^h 13^m 00^s.34$, $\delta = -19^\circ 31' 08''.7$, J2000) was
discovered in NGC~5018 on June 12.2 UT \citep[IAUC 7918,][]{Hutchings02}
and classified by ESC members as a Type Ia event on June 14.15 UT
\citep[IAUC 7919,][]{Riello02}. The layout of the manuscript is as
follows. Observations and data reduction are presented in Sect.~2.  We
describe the reddening estimate in Sect.~3. In Sect.~4  we analyze the
SN~2002dj optical and IR photometry comparing its  properties with those
of its kinematical twin, SN~2002bo. 
We also investigate whether they are representative of the HVG SNe
group. 
Sect.~5 contains the determination of the SN~2002dj absolute luminosity and a
characterization of the properties of its host galaxy. The optical and
IR spectroscopy is analyzed in Sect.~6 and Sect.~7, respectively.  The
expansion velocities are discussed in Sect.~8 and models of the early
and late spectra in Sect~9.  We conclude in Sect.~10 by examining
possible physical conditions in the HGV SNe ejecta that could 
explain part of their spectroscopic and photometric behavior.

\section[]{Observations and Data Reduction}

\begin{figure*} 
\centering
\includegraphics[width=0.48 \textwidth]{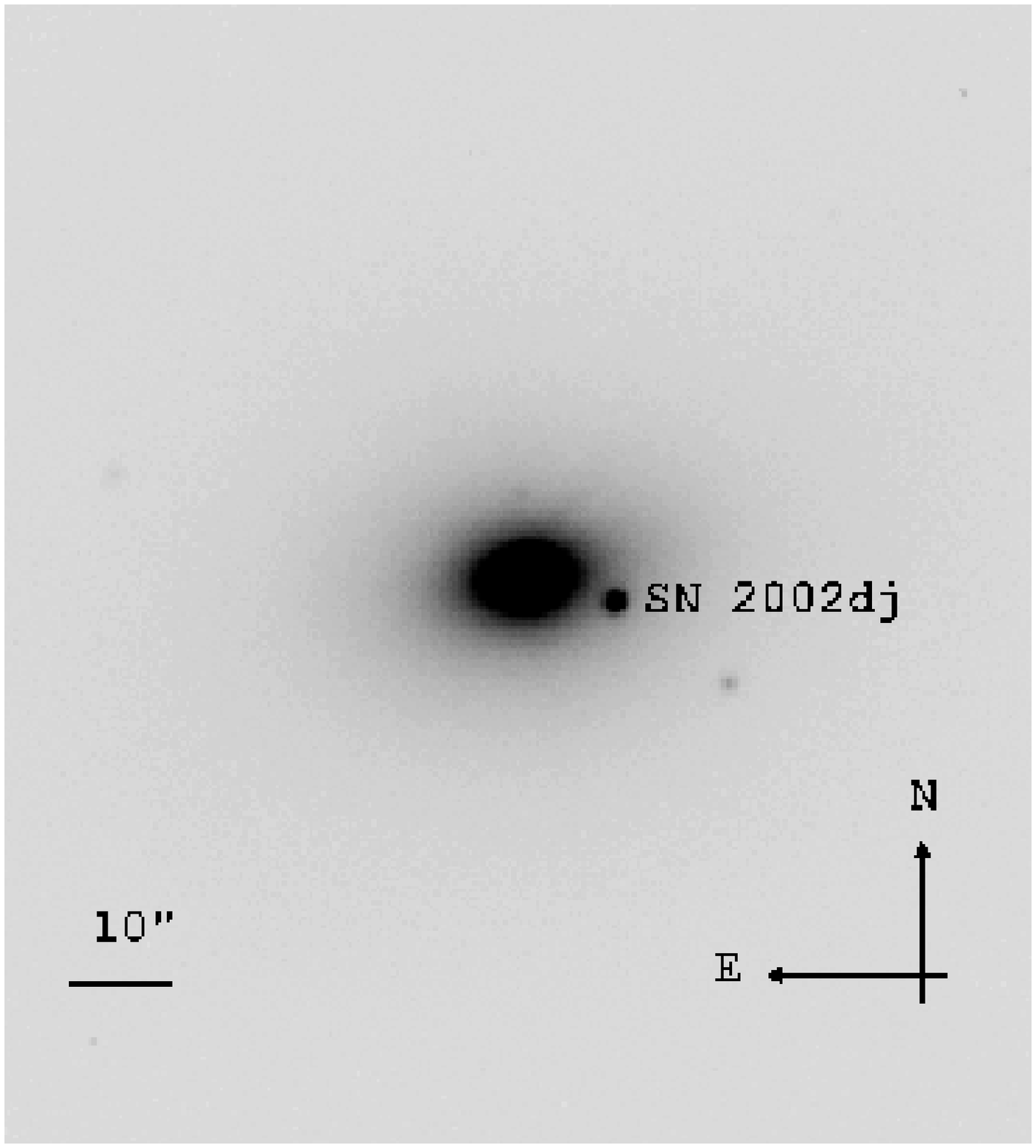}
\includegraphics[width=0.48 \textwidth]{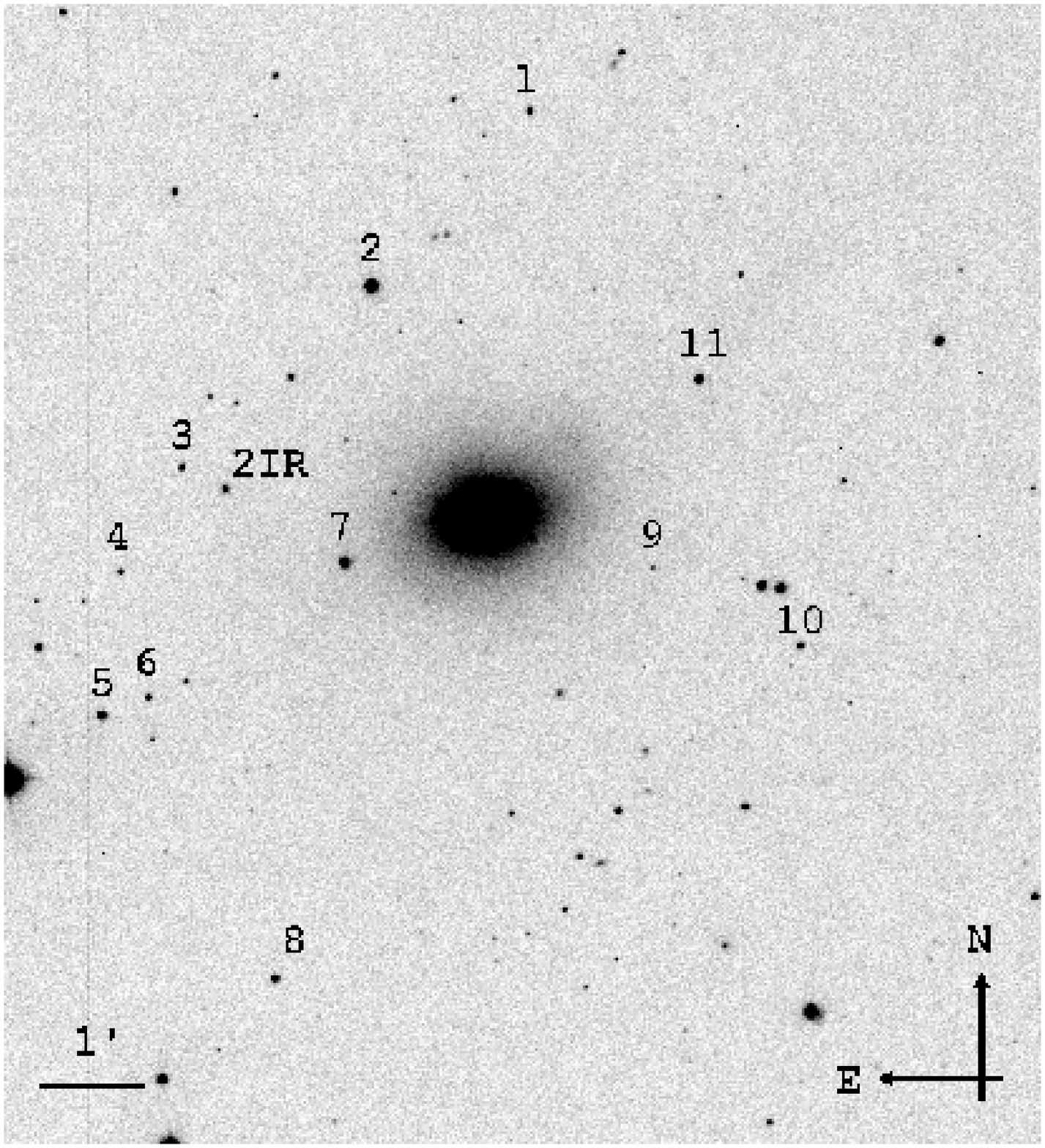}
\caption{Finding chart for SN~2002dj (left panel) and the local standard
stars (right panel) in a R-band exposure obtained on July 25 2002 with
the Danish~1.54m+DFOSC.} 
\label{fig2.1} 
\end{figure*}

\subsection{Instrument settings}
SN~2002dj was observed with seven different instruments in the optical
and two facilities in the near infrared.  Below we list the instruments
and their main characteristics:

\begin{itemize}

\item 0.9m telescope located at Cerro Tololo Inter-American Observatory
(CTIO) equipped with a CCD camera (2048$\times$2048, pixel size = 0.40
arcsec pixel$^{-1}$) and UBVRI Johnsons, Kron-Cousin standard filters.

\item 1.0m telescope located at CTIO equipped with A Novel
Double-Imaging CAMera (ANDICAM; Hawaii HgCdTe 1024$\times$1024, pixel
size = 0.137 arcsec pixel$^{-1}$) and JHKs standard filters.

\item European Southern Observatory (ESO) New Technology Telescope (NTT)
located at La Silla Observatory and equipped with the ESO Multi Mode
Instrument (EMMI) in RILD mode (2$\times$MIT/LL CCD 2048$\times$4096,
pixel size = 0.167 arcsec pixel$^{-1}$) and BVRI standard filters with
ESO identification numbers 605, 606, 608 and 610, respectively.

\item ESO-NTT equipped with the Son of ISAAC camera (SofI; Hawaii
HgCdTe 1024$\times$1024, pixel size = 0.29 arcsec pixel$^{-1}$) and JHKs
standard filters.

\item Danish 1.54m telescope located at La Silla Observatory equipped with
the Danish Faint Object Spectrograph and Camera (DFOSC; MAT/EEV CCD 44-82
2048$\times$2048, pixel size = 0.39 arcsec pixel$^{-1}$) and UBVRI
standard filters with ESO identification numbers 632, 450, 451, 452
and 425, respectively.

\item ESO-Kueyen Very Large Telescope (VLT) located at Paranal
Observatory equipped with the FOcal Reducer and low dispersion
Spectrograph (FORS1; 2$\times$E2V 2048$\times$4096 pixel size = 0.2 arcsec
pixel$^{-1}$) and BVRI standard filters with ESO identification
numbers +34, +35, +36 and +37, respectively.

\item Nordic Optical Telescope (NOT) located at Roque de los Muchachos
Observatory equipped with the Andaluc\'ia Faint Object Spectrograph and
Camera (ALFOSC; E2V 2048$\times$2048, pixel size = 0.19 arcsec
pixel$^{-1}$) and UBVR standard filters with NOT identification numbers
7, 74, 75 and 76, respectively, and an interference i filter with number
12.

\item Isaac Newton Group (ING) Jacobus Kapteyn Telescope (JKT) located
at Roque de los Muchachos  Observatory equipped with a CCD camera (SITe2
2048$\times$2048, pixel size = 0.33 arcsec pixel$^{-1}$) and UBVRI
standard filters with ING identification numbers 3, 27, 30, 37 and 44,
respectively.

\item ING-Isaac Newton Telescope (INT) located at Roque de los Muchachos
Observatory equipped with the Wide Field Camera (WFC; 4 thinned EEV
2048$\times$4096, pixel size = 0.33 arcsec pixel$^{-1}$) and
UBVRi(sloan) standard filters with ING identification numbers 204, 191,
192, 193 and 215, respectively.

\end{itemize}

\subsection{Optical and infrared photometry}

\begin{table*}
\begin{minipage}{110mm}
\caption{Magnitudes of the local photometric standard stars in the field of
SN~2002dj (Fig. \ref{fig2.1} right panel), obtained
during seven photometric nights with DFOSC, CTIO~0.9m, WFC and FORS1.}
\begin{tabular}{@{}lccccc}
\hline id & U & B & V & R & I \\
\hline
 1  & $-$               & 17.91 $\pm$  0.02 & 17.09 $\pm$  0.03 & 16.61 $\pm$  0.03 & 16.18 $\pm$  0.04 \\
 2  & 15.79 $\pm$  0.04 & 15.08 $\pm$  0.01 & 14.09 $\pm$  0.02 & 13.53 $\pm$  0.02 & 13.02 $\pm$  0.01 \\
 3  & $-$               & 18.11 $\pm$  0.02 & 17.33 $\pm$  0.04 & 16.88 $\pm$  0.03 & 16.43 $\pm$  0.02 \\
 4  & $-$               & 18.15 $\pm$  0.04 & 17.51 $\pm$  0.03 & 17.13 $\pm$  0.01 & 16.76 $\pm$  0.06 \\
 5  & 16.44 $\pm$  0.05 & 16.45 $\pm$  0.02 & 15.84 $\pm$  0.01 & 15.48 $\pm$  0.01 & 15.11 $\pm$  0.01 \\
 6  & $-$               & 18.59 $\pm$  0.02 & 17.52 $\pm$  0.05 & 16.90 $\pm$  0.01 & 16.31 $\pm$  0.01 \\
 7  & 15.05 $\pm$  0.03 & 15.08 $\pm$  0.01 & 14.62 $\pm$  0.02 & 14.33 $\pm$  0.01 & 14.04 $\pm$  0.03 \\
 8  & 18.01 $\pm$  0.08 & 17.60 $\pm$  0.01 & 16.57 $\pm$  0.01 & 15.96 $\pm$  0.02 & 15.38 $\pm$  0.01 \\
 9  & $-$               & 19.25 $\pm$  0.06 & 18.68 $\pm$  0.04 & 18.32 $\pm$  0.05 & 17.85 $\pm$  0.08 \\
10  & 17.31 $\pm$  0.07 & 17.28 $\pm$  0.03 & 16.62 $\pm$  0.01 & 16.22 $\pm$  0.04 & 15.85 $\pm$  0.03 \\
11  & 16.87 $\pm$  0.06 & 16.76 $\pm$  0.03 & 15.98 $\pm$  0.03 & 15.50 $\pm$  0.02 & 15.03 $\pm$  0.01 \\
\hline
\end{tabular}
\end{minipage}
\label{tab2.1}
\end{table*}

\begin{table*}
\begin{minipage}{170mm}
\caption{S-corrected optical photometry of SN~2002dj. The photometry obtained using the color equation are in Table~\ref{tabA.3}}
\begin{tabular}{@{}lcrcccccc}
\hline UT date & M.J.D. & Phase$^a$ & U & B & V & R & I & Instr. \\
\hline
13/06/2002$^p$ & 52439.1 & -10.9  & 16.07 $\pm$  0.04 & 16.06 $\pm$  0.02 & 15.85 $\pm$  0.02 & 15.66 $\pm$  0.02 & 15.90 $\pm$  0.03 & CTIO 0.9m \\
14/06/2002~ & 52439.9 & -10.1  &  $-$  & 15.64 $\pm$  0.02 & 15.49 $\pm$  0.02 & 15.27 $\pm$  0.02 & 15.46 $\pm$  0.02 & EMMI \\
27/06/2002$^p$ & 52452.9 &   2.9  & 14.30 $\pm$  0.03 & 14.36 $\pm$  0.02 & 14.14 $\pm$  0.02 & 14.10 $\pm$  0.02 & 14.46 $\pm$  0.02 & WFC \\
28/06/2002~ & 52453.9 &   3.9  &  $-$  & 14.41 $\pm$  0.02 & 14.16 $\pm$  0.02 & 14.12 $\pm$  0.02 & 14.48 $\pm$  0.02 & JKT \\
29/06/2002~ & 52455.0 &   5.0  & 14.41 $\pm$  0.04 & 14.46 $\pm$  0.02 & 14.20 $\pm$  0.02 & 14.15 $\pm$  0.03 & 14.51 $\pm$  0.03 & JKT \\
30/06/2002~ & 52456.0 &   6.0  &  $-$  &  $-$  &  $-$  & 14.24 $\pm$  0.02 &  $-$  & JKT \\
02/07/2002~ & 52458.0 &   8.0  &  $-$  & 14.66 $\pm$  0.09 & 14.31 $\pm$  0.02 & 14.31 $\pm$  0.02 & 14.61 $\pm$  0.02 & JKT \\
03/07/2002~ & 52458.1 &   8.1  & 14.86 $\pm$  0.04 & 14.67 $\pm$  0.02 & 14.27 $\pm$  0.02 & 14.30 $\pm$  0.02 & 14.66 $\pm$  0.05 & DFOSC \\
04/07/2002~ & 52460.0 &  10.0  & 15.02 $\pm$  0.04 & 14.81 $\pm$  0.02 & 14.36 $\pm$  0.02 & 14.46 $\pm$  0.02 & 14.85 $\pm$  0.05 & DFOSC \\
07/07/2002$^p$ & 52463.0 &  13.0  & 15.42 $\pm$  0.04 & 15.13 $\pm$  0.02 & 14.56 $\pm$  0.02 & 14.61 $\pm$  0.03 & 14.94 $\pm$  0.04 & DFOSC \\
09/07/2002~ & 52465.0 &  15.0  & 15.75 $\pm$  0.04 & 15.39 $\pm$  0.02 & 14.73 $\pm$  0.02 & 14.71 $\pm$  0.02 & 14.97 $\pm$  0.02 & DFOSC \\
11/07/2002~ & 52467.0 &  17.0  & 16.06 $\pm$  0.04 & 15.61 $\pm$  0.02 & 14.85 $\pm$  0.02 & 14.76 $\pm$  0.02 & 14.95 $\pm$  0.02 & DFOSC \\
17/07/2002$^p$ & 52473.0 &  23.0  &  $-$  &  $-$  & 15.12 $\pm$  0.02 & 14.86 $\pm$  0.02 & 14.81 $\pm$  0.02 & CTIO 0.9m \\
16/07/2002~ & 52473.1 &  23.1  & 16.75 $\pm$  0.04 & 16.29 $\pm$  0.02 & 15.13 $\pm$  0.02 & 14.83 $\pm$  0.02 & 14.89 $\pm$  0.02 & DFOSC \\
19/07/2002~ & 52476.0 &  26.0  & 16.99 $\pm$  0.04 & 16.54 $\pm$  0.02 & 15.40 $\pm$  0.02 & 15.05 $\pm$  0.02 & 14.85 $\pm$  0.02 & DFOSC \\
24/07/2002~ & 52479.9 &  29.9  &  $-$  & 16.89 $\pm$  0.02 & 15.58 $\pm$  0.02 & 15.14 $\pm$  0.02 & 14.83 $\pm$  0.02 & ALFOSC \\
25/07/2002~ & 52481.0 &  31.0  & 17.44 $\pm$  0.05 & 16.91 $\pm$  0.02 & 15.59 $\pm$  0.02 & 15.18 $\pm$  0.02 & 14.88 $\pm$  0.02 & DFOSC \\
29/07/2002~ & 52484.9 &  34.9  & 17.59 $\pm$  0.15 &  $-$  &  $-$  &  $-$  &  $-$  & ALFOSC \\
08/08/2002$^p$ & 52495.0 &  45.0  &  $-$  & 17.41 $\pm$  0.04 & 16.25 $\pm$  0.02 & 15.95 $\pm$  0.03 & 15.77 $\pm$  0.06 & CTIO 0.9m \\
31/08/2002~ & 52518.0 &  68.0  & 18.43 $\pm$  0.09 &  $-$  & 16.83 $\pm$  0.02 &  $-$  &  $-$  & DFOSC \\
02/09/2002~ & 52520.0 &  70.0  &  $-$  &  $-$  & 16.87 $\pm$  0.02 &  $-$  &  $-$  & DFOSC \\
06/09/2002~ & 52524.0 &  74.0  &  $-$  & 17.75 $\pm$  0.19 &  $-$  & 17.06 $\pm$  0.06 & 17.01 $\pm$  0.48 & DFOSC \\
25/03/2003$^p$ & 52724.3 & 274.3  &  $-$  & 21.00 $\pm$  0.04 & 20.71 $\pm$  0.03 & 21.39 $\pm$  0.06 & 20.78 $\pm$  0.09 & FORS1 \\
22/05/2004$^p$ & 53147.2 & 697.2  &  $-$  & $>$24.7$^b$ & $>$23.4$^b$ &   $-$  &  $-$  & FORS1 \\
\hline
\end{tabular}
\\
$^a$ Counted since the time of the $B$ maximum brightness M.J.D.=52450 $\pm$ 0.7\\
$^b$ Upper limit\\
$^p$ Photometric night\\
\end{minipage}
\label{tab2.2}
\end{table*}

\begin{table}
\caption{Magnitudes for the local IR photometric sequence in the field of
SN~2002dj (Fig. \ref{fig2.1} right panel). The data were obtained during three
photometric nights with SofI. }
\begin{tabular}{@{}lccc}
\hline id & J & H & Ks  \\
\hline
 2  & 12.23 $\pm$  0.03 & 11.75 $\pm$  0.03 & 11.63 $\pm$  0.03 \\
 2 IR  & 14.80 $\pm$  0.03 & 14.18 $\pm$  0.03 & 13.95 $\pm$  0.04 \\
 7  & 13.63 $\pm$  0.04 & 13.42 $\pm$  0.04 & 13.38 $\pm$  0.04 \\
 11  & 14.32 $\pm$  0.04 & 13.89 $\pm$  0.04 & 13.76 $\pm$  0.04 \\
\hline
\end{tabular}

\label{tab2.3}
\end{table}

For the optical photometric observations basic data reduction (bias
and flat-field correction) was performed using standard routines in
IRAF\footnote{IRAF is distributed by the National Optical Astronomy
Observatories, which are operated by the Association of Universities
for Research in Astronomy, Inc, under contract to the National Science
Foundation.}. In Table~1 are reported the magnitudes of the local photometric
sequence identified in Fig.~\ref{fig2.1}, that were calibrated against Landolt standard
stars \citep{Landolt92} in the seven photometric nights marked in Table~2. SN photometry was performed using the PSF
fitting technique. Only in the $B$ and $V$ FORS1 images obtained on
March 25, 2003, was it possible to remove the galaxy. For the other
instruments the lack of a suitable image without the SN has prevented
us from using the template subtraction technique. We note that the
FORS1 $B$ and $V$ band frames on which the SN is still relatively
bright yield magnitudes in agreement to the ones derived from the
template subtracted images. This gives us confidence that the PSF
photometry performed around maximum, when the SN signal-to-noise ratio
is high, was not significantly contaminated by the host galaxy
background.

The near-IR data reductions were also performed using standard IRAF
routines. Dark and flat-field corrections were applied to the
scientific frames. For each night a sky background image was created
by taking the median of the dithered science frames. For the ANDICAM
images this approach lead to an imperfect removal of the galaxy light.
Nevertheless, the good agreement between the SofI and ANDICAM
photometry gives us confidence that the small galaxy residuals did not
significantly bias the ANDICAM magnitudes.  The sky background image
was then subtracted from each single frame. For SofI, an illumination
correction was also applied to all images.  Four stars close to the SN
position were calibrated in the $JHKs$ bands using standard stars from
\citet{Persson98} during three photometric nights marked in
Table~\ref{tab2.4}. The resulting magnitudes are listed in
Table~\ref{tab2.3}. They agree  well with those reported in the 2MASS
catalog with $J_{seq}-J_{2MASS}$=0.01 $\pm$ 0.01,
$H_{seq}-H_{2MASS}$=0.00 $\pm$ 0.01 and $Ks_{seq}-Ks_{2MASS}=0.01 \pm
0.01$.  The final SN calibration in all bands was performed using the
S-correction technique (see Appendix for details). Thanks to the very
good spectroscopic coverage we could compute the S-terms using only
the spectra of SN~2002dj for the $B$, $V$, $R$, $I$ bands,  while for
the $U$, $J$, $H$, $Ks$, the SN~2002dj spectra were complemented with
those of SN~2005cf \citep{Garavini07} and SN~2002bo \citep{Benetti04}.

\begin{table*}

\begin{minipage}{110mm}
\caption{S-corrected IR photometry of SN~2002dj. The photometry obtained using the color equation are in Table~\ref{tabA.4}}
\label{tab2.4}
\begin{tabular}{@{}lcrcccc}
\hline UT date & M.J.D. & Phase$^a$ & J & H & Ks & Instr. \\
\hline
13/06/2002~ & 52439.0 & -11.0  & 15.63 $\pm$  0.06 & 15.64 $\pm$  0.09 & 15.65 $\pm$  0.13 & ANDICAM \\
14/06/2002~ & 52440.0 & -10.0  & 15.27 $\pm$  0.04 & 15.31 $\pm$  0.06 & 15.25 $\pm$  0.09 & ANDICAM \\
15/06/2002~ & 52441.1 &  -8.9  & 14.97 $\pm$  0.03 & 15.13 $\pm$  0.03 & 15.11 $\pm$  0.03 & SoFi \\
19/06/2002$^p$ & 52444.1 &  -5.9  & 14.63 $\pm$  0.04 & 14.84 $\pm$  0.03 & 14.71 $\pm$  0.03 & SoFi \\
20/06/2002~ & 52446.0 &  -4.0  & 14.58 $\pm$  0.11 & 14.80 $\pm$  0.07 & 14.66 $\pm$  0.08 & ANDICAM \\
24/06/2002~ & 52450.0 &  -0.0  & 14.60 $\pm$  0.03 & 14.84 $\pm$  0.06 & 14.52 $\pm$  0.10 & ANDICAM \\
24/06/2002$^p$ & 52450.0 &  -0.0  & 14.62 $\pm$  0.03 & 14.83 $\pm$  0.03 & 14.54 $\pm$  0.03 & SoFi \\
27/06/2002~ & 52453.1 &   3.1  & 14.86 $\pm$  0.05 & 14.90 $\pm$  0.04 & 14.68 $\pm$  0.09 & ANDICAM \\
30/06/2002~ & 52456.0 &   6.0  & 15.30 $\pm$  0.13 & 15.09 $\pm$  0.10 & 14.88 $\pm$  0.20 & ANDICAM \\
08/07/2002~ & 52464.0 &  14.0  & 16.30 $\pm$  0.09 & 15.05 $\pm$  0.05 & 15.04 $\pm$  0.09 & ANDICAM \\
11/07/2002~ & 52467.0 &  17.0  & 16.26 $\pm$  0.05 & 14.97 $\pm$  0.03 & 14.90 $\pm$  0.06 & ANDICAM \\
11/07/2002$^p$ & 52467.0 &  17.0  & 16.20 $\pm$  0.03 & 14.95 $\pm$  0.03 & 14.83 $\pm$  0.03 & SoFi \\
14/07/2002~ & 52470.0 &  20.0  & 16.08 $\pm$  0.13 & 14.81 $\pm$  0.05 & 14.75 $\pm$  0.08 & ANDICAM \\
17/07/2002~ & 52473.0 &  23.0  & 15.86 $\pm$  0.04 & 14.75 $\pm$  0.03 & 14.65 $\pm$  0.05 & ANDICAM \\
25/07/2002~ & 52481.0 &  31.0  & 15.57 $\pm$  0.07 & 14.79 $\pm$  0.05 & 14.81 $\pm$  0.09 & ANDICAM \\
28/07/2002~ & 52484.0 &  34.0  & 15.63 $\pm$  0.05 & 15.06 $\pm$  0.04 & 15.12 $\pm$  0.07 & ANDICAM \\
31/07/2002~ & 52487.0 &  37.0  & 15.89 $\pm$  0.03 & 15.23 $\pm$  0.07 & 15.43 $\pm$  0.29 & ANDICAM \\
07/08/2002~ & 52494.0 &  44.0  & 16.58 $\pm$  0.14 & 15.61 $\pm$  0.09 & 15.65 $\pm$  0.35 & ANDICAM \\
10/08/2002~ & 52497.0 &  47.0  & 16.68 $\pm$  0.10 & 15.73 $\pm$  0.06 & 15.93 $\pm$  0.16 & ANDICAM \\
13/08/2002~ & 52500.0 &  50.0  & 16.89 $\pm$  0.12 & 15.85 $\pm$  0.06 & 16.04 $\pm$  0.15 & ANDICAM \\
30/08/2002~ & 52517.0 &  67.0  & 18.20 $\pm$  0.38 & 16.66 $\pm$  0.03 & 16.84 $\pm$  0.03 & SoFi \\
\hline
\end{tabular}
$^a$ Counted since the time of the $B$ maximum brightness M.J.D.=52450 $\pm$ 0.7\\
$^p$ Photometric night\\
\end{minipage}
\end{table*}

\subsection{Optical and infrared spectroscopy}

Optical spectra were reduced using the {\it ctioslit} package in IRAF.
Optimal extraction was usually obtained by weighting the signal
according to the intensity profile along the slit. Sky subtraction was
carried out by fitting a low order polynomial on either side of the SN
spectrum and the wavelength solution was determined from arc spectra.
The wavelength calibration was checked against bright night-sky
emission lines. Flux calibration was performed by means of
spectrophotometric standard stars \citep{Hamuy92,Hamuy94} and checked
against the photometry. When discrepancies occurred, the flux of the
spectrum was scaled to match the photometry. In nights with good
observing conditions the agreement with the photometry was found to be
within 10\%.

The near-IR spectra were reduced using standard procedures in the
FIGARO 4 environment \citep{Shortridge95}. Wavelength calibration was
carried out using Xe arc spectra and the accuracy of the solution was
checked using OH sky lines. The flux calibration was secured with
respect to near-IR solar-analog standard stars observed close in time
and airmass to the SN to minimize variations in the atmospheric
absorptions \citep{Maiolino96}.  The SN spectrum was divided by the
standard stars spectrum to remove the strong telluric IR features. The
resulting spectrum was then multiplied by the solar spectrum to
eliminate the intrinsic features introduced by the solar-type star.
Like in the optical, the flux calibration was checked against the
photometry and, if necessary, scaled to match the latter.

\section{Interstellar extinction}

Presence of dust in NGC~5018 has been claimed in several studies. The
map produced by \citet{Fort86} shows a dust lane embedded in the
galaxy. Subsequently, \citet{Carollo94} and \citet{Goudfrooij94a}
detected patchy dust in the galaxy core.  None of those maps shows
evidence of dust at the location of SN~2002dj. Our deep images taken
with the VLT add details to the \citet{Carollo94} colour maps, but
again do not show the presence of dust at, or close to, the SN
position. 
SN spectra can be used to measure  the equivalent width ($EW$) of the
Na~I~D lines as a proxy to quantify absorption. The Galactic colour
excess we obtain using one of the two linear relations reported in
\citet{Turatto03} ($E(B-V)=0.16 \times EW$ yielding
$E(B-V)_{Galactic}=0.11 \pm 0.03$ from $EW=0.72 \pm 0.19$ \AA) is in
very good agreement with that reported by \citet{Schlegel98}
($E(B-V)_{Galactic}=0.096$) for the SN~2002dj line of sight.  No
absorption line of interstellar Na~I~D has been detected from the host
galaxy. Since SN~2002dj was observed both in the optical and in the
near IR, we can apply several methods to estimate its extinction from
the observed colour. The results are reported in Table~\ref{tab3.4}.
We note that all methods involving $B-V$ point to an $E(B-V)$ greater
than that due to Milky Way absorption alone, indicating a small but
not negligible dust extinction towards SN~2002dj in the host galaxy,
while the methods based on $V-IR$ colours provide a negligible colour
excess for the host galaxy. A similar apparent inconsistency occurred
in the case of SN~2002bo, where the reddening derived from $B-V$
colour seems to be larger than that derived through detailed synthetic
spectra modeling \citep{Stehle05} and, as in the case of SN~2002dj, it
is significantly larger than that derived from the $V-IR$ colours. The
peculiar colours of SN~2002bo were already noted by \citet{Benetti04}
and \citet{Krisciunas04b}.

With the aim to check whether HVG SNe could in general have
intrinsically different colours which could bias their reddening
estimate, we examined the other HVG SNe reported in the literature for
which multi-epoch spectra show a persistent high expansion velocity
(see Table \ref{tab5.3})\footnote{Although SN~2002er was classified as
HVG SNe \citep{Benetti05} it was not included in the table because of
its lower expansion velocities. This SN could indeed represents a
transition object between HVG SNe and LVG SNe \citep{Tanaka08}.}.
Unfortunately, all objects except SN~2002bf and SN~2004dt, show signs
of absorption in their host galaxies (i.e. Na~I~D lines). The latter two
SNe are also not useful for this purpose because the photometric
follow-up for SN~2002bf started $\sim$ at +6 days and definitive
photometry is not yet available for SN~2004dt. Hence, for all objects
in Table \ref{tab5.3} the intrinsic colour is  rather uncertain. To
investigate the possible peculiar colours for HVG SNe we compared the
the scatter between the colour excess derived using
the $B-V$ and $V-I$ colours among LVG SNe with the systematic difference
we found for our sample of HVG SNe.

To determine the difference in extinction measure for LVG SNe we used
Table 2 of \citet{Phillips99} and obtained
$E(B-V)_{B-V}$-$E(B-V)_{V-I}=-0.01 \pm 0.08$. This value was derived
by selecting SNe with $0.95 < \Delta m_{15} < 1.60$ in order to avoid
colour peculiarity not accounted by the \citet{Phillips99} relation
\citep{Garnavich04}. 
For HVG SNe in addition to SN~2002bo and SN~2002dj, only in the case
of SN~1997bp, {\it BVRI} photometry  \citep{Altavilla04,Jha06} was suitable to compute the colours at maximum. 
For these three objects we find $E(B-V)_{B-V}$-$E(B-V)_{V-I}=0.17 \pm
0.03$. The systematic difference in reddening depending on the
reference color in HVG SNe is hence nearly $2\sigma$ the scatter
observed in LVG SNe hinting at intrinsic colour differences.
Anyway, the colour evolution among HVG SNe shows significant variations (see
Fig.~\ref{fig4.5}) making it difficult to establish an overall
difference with LVG SNe. 

\begin{table}

\caption{SN~2002dj total (Galactic + host galaxy) reddening from different methods.}
\hspace{15pt}\\
\begin{tabular}{@{}lclc}
 \hline
  Method & $E(B-V)$ & Reference \\
\hline
$B_{max}-V_{max}$ & 0.22 $\pm$ 0.06 & \citet{Phillips99}  \\
$(B-V)_{max}$ & 0.24 $\pm$ 0.06 & \citet{Altavilla04}  \\
$V_{max}-I_{max}$ & 0.10 $\pm$ 0.06 &  \citet{Phillips99}  \\
$(B-V)_{CMAGIC}$ & 0.32 $\pm$ 0.06 &  \citet{Wang03}  \\
$(B-V)_{Jha}$ & 0.29 $\pm$ 0.06 &  \citet{Jha06}  \\
$(B-V)_{tail}$ & 0.27 $\pm$ 0.14 &  \citet{Phillips99} \\
$(B-V)_{Wang}$ &  0.21 $\pm$ 0.04 &  \citet{Wang05} \\
$(B-V)_{V-H}$ & 0.05 $\pm$ 0.04 & \citet{Krisciunas04a} \\
$(B-V)_{V-K}$ & 0.10 $\pm$ 0.06 & \citet{Krisciunas04a} \\
\hline
\end{tabular}
\label{tab3.4}
\end{table} 

\section{Optical and Infrared photometry}
\begin{figure*}
\includegraphics[width=175mm]{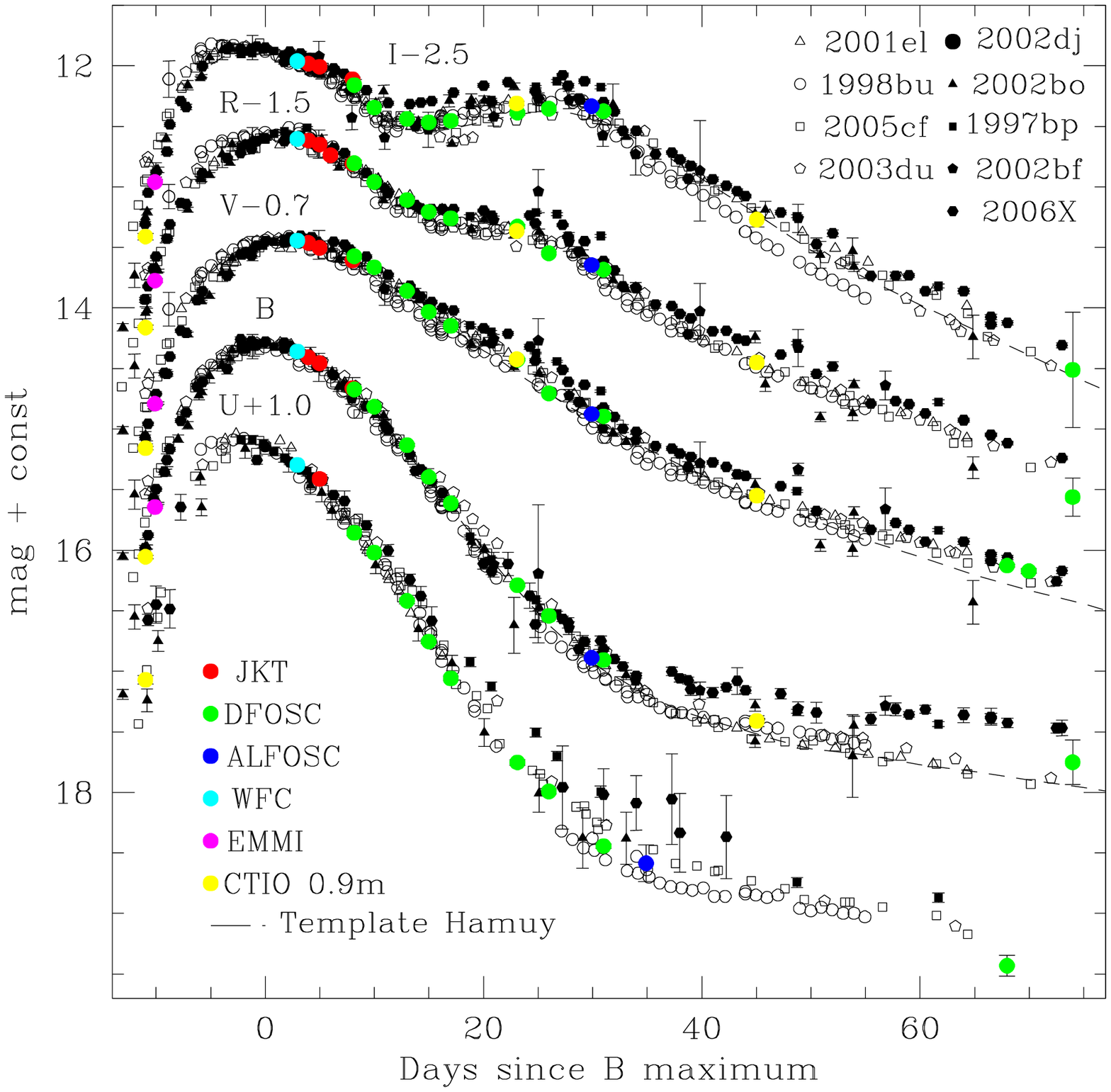}
\caption{$UBVRI$ light curves of SN~2002dj. Our data are shown with
filled circles using different colours for different instruments.  For
comparison the light curves of other SNe as well as  the template for
$BVI$ $\Delta m_{15}=1.11$ by \citet{Hamuy96} (dashed lines) are
displayed. HVG SNe are indicated by filled symbols: SN~2002bo
\citep{Benetti04,Krisciunas04b}, SN~1997bp \citep{Altavilla04, Jha06},
SN~2002bf \citep{Leonard05}, SN~2006X \citep{Wang07a}, while LVG SNe
are drawn with open symbols: SN~1998bu \citep{Jha99}, SN~2001el
\citep{Krisciunas03}, SN~2005cf \citep{Pastorello07} and SN~2003du
\citep{Stanishev07}. To facilitate the comparison the light curves of
the different SNe were normalised to match at maximum (between $-$5 to
$+$5 days).}
\label{fig4.1}
\end{figure*}

\begin{figure}
\centering
\includegraphics[width=84mm]{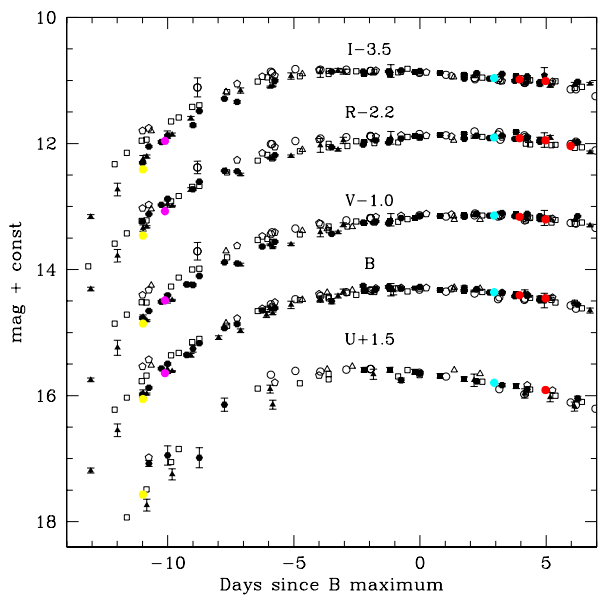}
\caption{Early phase $UBVRI$ light curves of SN~2002dj. The symbols, SNe and vertical shifts are the same as reported in Fig.~\ref{fig4.1}.
}
\label{fig4.4}
\end{figure}

SN~2002dj was observed in $UBVRI$ from $-$11 to +697 days and from
$-$11 to +68 days in $JHKs$. The corresponding light curves are shown
in Figs.~\ref{fig4.1},~\ref{fig4.4},~\ref{fig4.2},~\ref{fig4.3}, while
colour curves are displayed in Figs.~\ref{fig4.5} and \ref{fig4.6}.

\subsection{Light curves}

\begin{figure}
\centering
\includegraphics[width=84mm]{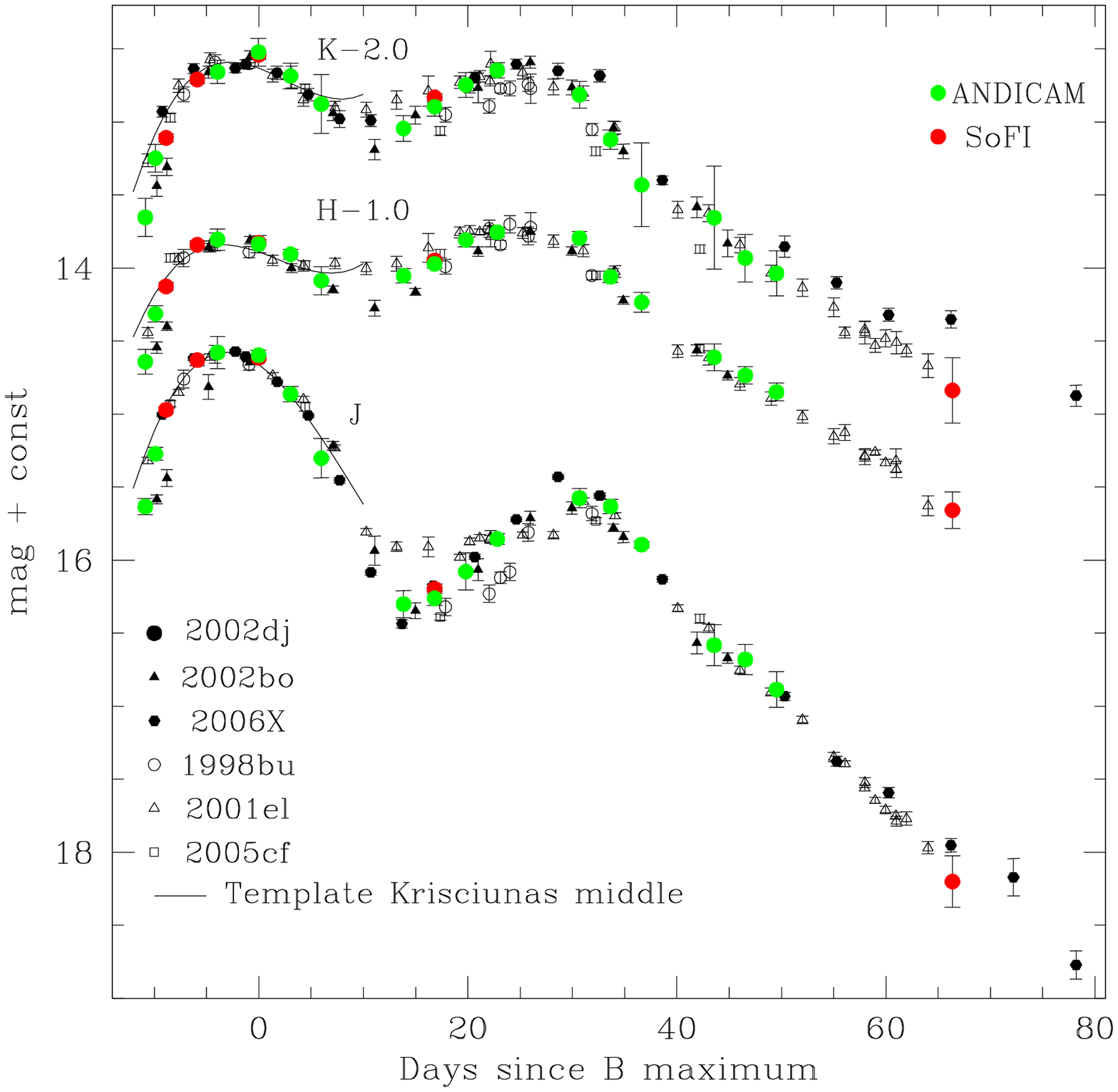}
\caption{$JHK$ light curves of SN~2002dj, different colours refer to
different instruments.  For comparison the light curves of SN~2002bo
\citep{Krisciunas04b}, SN~2006X \citep{Wang07a}, SN~1998bu
\citep{Hernandez00}, SN~2001el \citep{Krisciunas03}, SN~2005cf
\citep{Pastorello07} and the mid range decliner templates from
\citet{Krisciunas04a} (solid line) are displayed. The light curves of
different SNe were vertically shifted in order to match them at
maximum.}
\label{fig4.2}
\end{figure}

In Fig.~\ref{fig4.4} the $BVI$ and, to a lesser degree $R$, HVG SNe (filled symbols; SN~2002dj, SN~2002bo
and SN~2006X) show a faster 
rise to maximum if compared to LVG
SNe (empty symbols; represented by SN~2001el, SN~2003du and
SN~2005cf). On the other hand we note that SN~2006X and SN~1997bp peak
later in $VRI$ ($\sim$+2 days) than the LVG average population (see
histograms in \citet{Contardo02}). Therefore the fast rise in $VRI$ of
the HVG objects could be induced by a peculiar behaviour of the $B$
light curve that historically is used to set the phase of the SN~Ia.
Comparing the light curves of  SN~2006X (SN~1997bp does not have
observations close to maximum) with those of SN~2005cf and leaving the
$B$ maximum time as a free parameter in the fit, we found a phase
shift of $-$0.9 days for the $V$ and $R$ bands and $-$0.5 for the $I$.
The latter results  reduce the possibility that the early time
behaviour of HVG SNe could simply be a phase mis-calibration.  This is
also confirmed by the IR observation of SN~2002dj, SN~2002bo and
SN~2006X which show that those SNe reach the peak at the same epoch as
the \citet{Krisciunas04a} templates (Fig.~\ref{fig4.2}).  For the $U$
band the situation is more complex. The data are more scattered and
systematic errors due to standard filter mis-matching could play an
important role. Therefore a photometrically more homogeneous data set
is necessary to confirm the trends observed in the other filters.

Another feature visible in Fig.~\ref{fig4.1} is that the HVG
SN~1997bp, SN~2002bf\footnote{The light curves of SN~2002bf have been
K-corrected to the host galaxy rest frame velocity of SN~2002dj}
 and SN~2006X show more pronounced inflections
around +25 days in $V$ and $R$. 
In the case of SN~2002bf the large photometric errors associated with
the corresponding points demand caution, while for
 SN~2006X, at least part of this effect could be due to its high
extinction shifting the effective wavelength of the passband to redder
\citep{EliasRosa06,Wang07a}. The absorption suffered by SN~1997bp is low
and the photometric errors small, hence this feature must be
intrinsic. Finally, we note that SN~1997bp, SN~2002bf and SN~2006X are
the highest velocity SNe~Ia in our sample, with velocities about 5000
km s$^{-1}$ higher than those of SN~2002dj and SN~2002bo
(Fig.~\ref{fig8.1}). This suggest that photometric features could be
correlated with the kinematics of the ejecta.

Since, as previously mentioned, at early epochs the light curves of
HVG SNe are different from those of LVG SNe, we decided to constrain
the epoch of the SN~2002dj $B$ maximum brightness, hereafter
T($B_{max}$), using only the $B$ light curve of SN~2002bo. In fact, it
is not viable to determine the maximum brightness directly from the
SN~2002dj light curve because it happens to fall in a gap of 12 days
with no photometric observations. Having constrained T($B_{max}$),
(M.J.D.=$52450.0 \pm 0.7$)  we used the light curves of SNe with
similar behavior around maximum to estimate the SN~2002dj peak
magnitude in $UBVRI$ (Table~\ref{tab5.1}). The values of the light
curve parameters $\Delta m_{15}$ and stretch factor $s$
\citep{Perlmutter97a} are reported in Table~\ref{tab5.2}.  The stretch
factor was obtained by fitting the SN~2002dj $B$ light curve to the
\citet{Leibundgut88} template complemented by \citet{Goldhaber01}.  We
note that the values of $\Delta m_{15}$ and $s$, satisfy the relations
reported by \citet{Perlmutter97b} and \citet{Altavilla04} reasonably
well.

As in the optical, in the near infrared bands both SN~2002dj and
SN~2002bo show a faster rise to maximum brightness than SN~2001el  and
the \citet{Krisciunas04a} ``mid-range'' decliner IR templates.
SN~2006X follows the templates closely, although 
very early time observations are missing.

At epochs $>$+35 days in $I$, $R$ and $V$ HVG SNe are perhaps slightly
brighter, but the points show large spread, covering possible subtle
systematic differences. In $B$ HVG SNe are clearly brighter than LVG
SNe. Interestingly, as in the case of the inflections occurring around
+25 days, the brightest objects among HVG SNe are those displaying the
highest expansion velocities. \citet{Wang07a} already  pointed out
that the decline rates $\gamma$ between +40 and +100 days (Pskovskii
1984) of the HVG SN~1984A, SN~2002bo and SN~2006X are smaller  than
for ``normal'' SNe~Ia (i.e. $\gamma$ = 1.40 $\pm$ 0.10 mag 100
d$^{-1}$), being 1.14 $\pm$ 0.06, 1.17 $\pm$ 0.10 and 0.92 $\pm$ 0.05,
respectively.  Fitting the three $B$ observations of SN~2002bf at
phase $>$ +40.0 days we obtained $\gamma$ = 0.90 $\pm$ 0.60. The large
error is due to the noisy photometry.  For SN~2002dj, considering the
only two $B$ observations in the phase interval, we derived $\gamma$
=1.17 $\pm$ 0.13. Also for SN~1997bp we have only two points from
which yield $\gamma$ =0.83 $\pm$ 0.18.  Therefore, although with
limited statistical strength, we confirm the finding of
\citet{Wang07a}. In $U$ the fastest expanding SN~1997bp and SN~2006X
seem to follow the $B$ band trend, while SN~2002dj and SN~2002bo are
closer to the evolution of SN~2005cf and SN~1998bu.

\begin{figure}
\centering
\includegraphics[width=84mm]{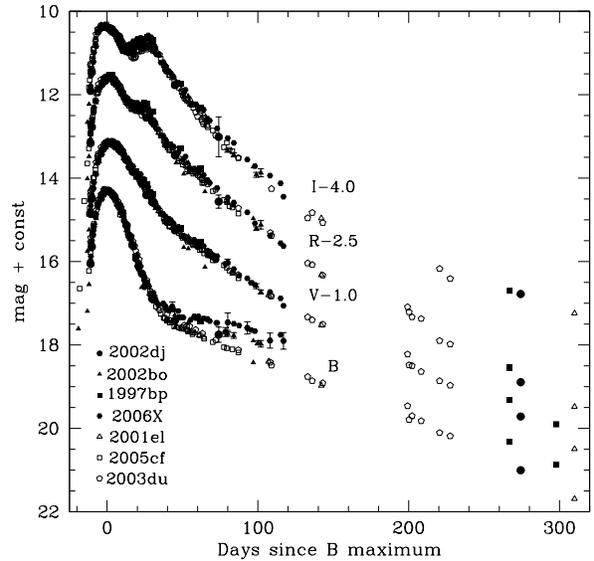}
\caption{Same as Fig.~\ref{fig4.1} showing the late phase $BVRI$ light
curves of SN~2002dj.}
\label{fig4.3}
\end{figure}

At late phases SN~2002dj, although an HVG SN resembles the
behavior of LVG SNe. At the latest epoch for which we obtained
measurable data (+274 days), SN~2002dj appears in fact in all bands close
to the light curve of SN~2001el \citep[][see
Fig.~\ref{fig4.3}]{Stritzinger07}. SN~1997bp is instead
$\sim$0.5 mag and $\sim$0.3 brighter in $B$ and $V$,
respectively. Its late time decline rate is similar to the other
SNe, which argues against a possible contribution of a light
echo.
With the aim to further investigate (or exclude) the presence of
a light echo, $B$ and $V$ imaging of SN~2002dj was performed with
FORS1 on April 22, 2004, i.e. 697.2 days after $B$ maximum light.
Nothing is visible at the position of the SN and through
re-detection of artificial stars, we estimated 3$\sigma$ upper
limits of $B$ $\sim$24.7 and $V$$\sim$23.4.

\subsection{Colour curves}

\begin{figure}

\centering
\includegraphics[width=84mm]{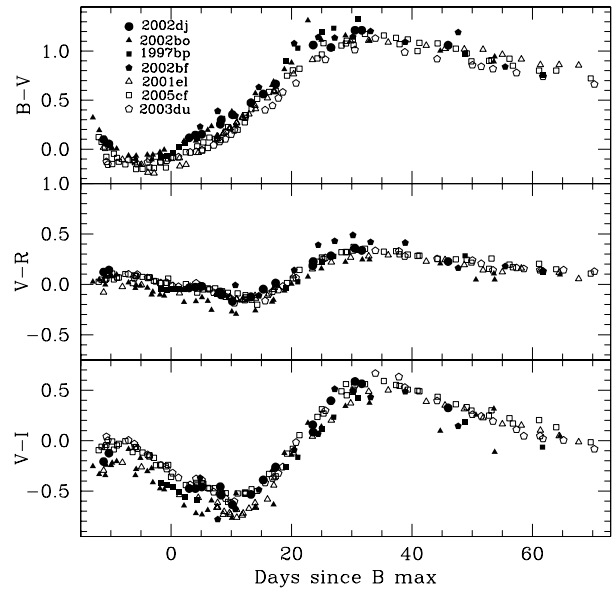}
\caption{De-reddened $(B-V)_0$, $(V-R)_0$ and $(V-I)_0$ colour
curves of SN~2002dj.  
}
\label{fig4.5} 
\end{figure}

The de-reddened colour curves of SN~2002dj are compared in
Fig.~\ref{fig4.5} with those of SN~1997bp, SN~2002bo, SN~2002bf,
SN~2001el, SN~2003du and SN~2005cf. For HVG SNe the reddening was
removed using the values of $E(B-V)$ reported in
Table~\ref{tab5.3} and assuming $R_V$=3.1. Because of their large
reddening, SN~2002bo and SN~2006X are not considered in the HVG
{\it vs.} LVG comparison.

In the $(B-V)_0$ colour, at least until +30 days, HVG SNe form a
redder sequence than LVG SNe. Afterward the two groups mix due to the HVG slower decline rate in the $B$ band.  Note
that the latter poses problems if one
wishes to apply the Lira relation \citep{Phillips99} to estimate
the reddening of HVG \citep[see also][]{Wang07a}. Therefore, both
maximum light and tail $B-V$ colours could bias the reddening
estimation of these objects. In $(V-R)_0$ there is not a clear
separation between HVG and LVG SNe, while in $(V-I)_0$ HVG SNe
seem to form a bluer sequence with respect to LVG SNe, but the
significance of the differences is reduced by the large
calibration uncertainties of the $I$ band
\citep[see][]{Pignata04}.

The $(V-R)_0$ and $(V-I)_0$ colours of SN~2002bo are
clearly bluer then those of SN~2002dj. This reflects the lack of
flux in the red part of the SN~2002bo spectrum noticeable in the
comparison with SN~2002dj (see section 6.3).

\begin{figure}

\centering
\includegraphics[width=84mm]{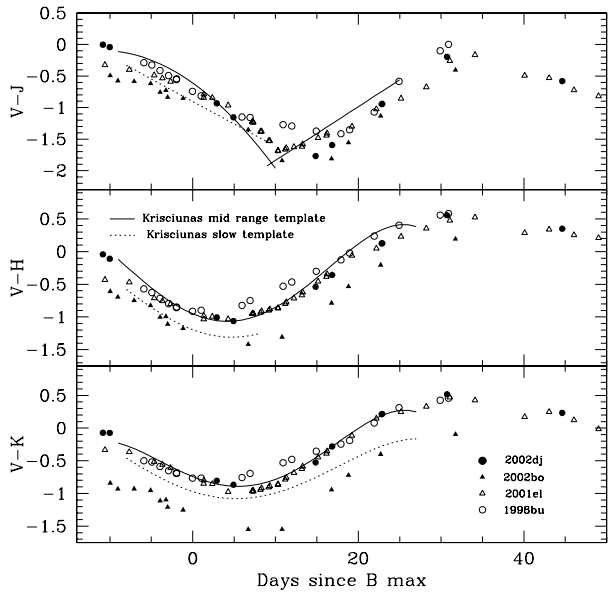}

\caption{De-reddened $(V-J)_0$, $(V-H)_0$ and $(V-K)_0$ colour curves
of SN~2002dj. 
}
\label{fig4.6}
\end{figure}

The $(V-J)_0$, $(V-H)_0$ and $(V-K)_0$ colours curves of
SN~2002dj (Fig.~\ref{fig4.6}) follow the \citet{Krisciunas04a}
``mid range'' templates reinforcing the idea of negligible host
galaxy reddening. As in the case of the $(V-R)_0$ and $(V-I)_0$
colours, SN~2002bo is bluer at all epochs. The difference to
SN~2002dj increases toward redder bands, suggesting that the lack
of flux starting around 6500 \AA~continues monotonically towards
longer wavelengths.

\section{Absolute luminosity}

The Virgo infall model of \citet{Kraan-Korteweg86} yields a
distance to relative Virgo of 2.52 for NGC~5018. Assuming a Virgo
distance of 15.3 Mpc \citep{Freedman01} we obtain 38.55 Mpc
($\mu=32.93$) for the host galaxy of SN~2002dj.  The Nearby
Galaxies Catalog \citep{Tully88} reports a larger distance
(40.9~Mpc, $\mu$=33.1), while \citet{Hilker96} using the globular
cluster luminosity function derived a distance modulus ranging
between 32.8 and 34.0.
For a consistent comparison to SN~2002bo we decided to assume
38.55 Mpc as the best distance estimate for SN~2002dj. 
With this distance and the reddening estimate we derived the
corresponding absolute magnitudes in $UBVRIJHKs$. The values are
reported in Table~\ref{tab5.1}, along with those obtained from
the $\Delta m_{15}$ absolute magnitude relations of
\citet{Prieto06} for $BVRI$ and those of \citet{Krisciunas04} for
the $JHKs$.  The absolute magnitudes in all the bands are in
agreement within 2$\sigma$ with the values predicted by the
previously mentioned relations, although SN~2002dj appears
slightly underluminous.

\begin{table}
\caption{Peak  magnitudes of SN~2002dj}
\hspace{15pt}\\
\begin{tabular}{@{}ccccc}

 \hline
  Filter & m(obs)$^a$ & M(cor)$^b$ & M(ave)$^c$ \\
  \hline
$B$ & 14.30 $\pm$ 0.04 & -19.03 $\pm$ 0.23 & -19.34 $\pm$ 0.02 \\
$V$ & 14.15 $\pm$ 0.04 & -19.08 $\pm$ 0.19 & -19.26 $\pm$ 0.02 \\
$R$ & 14.10 $\pm$ 0.05 & -19.06 $\pm$ 0.17 & -19.26 $\pm$ 0.02 \\
$I$ & 14.35 $\pm$ 0.06 & -18.72 $\pm$ 0.14 & -19.00 $\pm$ 0.02 \\
$J$ & 14.56 $\pm$ 0.02 & -18.46 $\pm$ 0.11 & -18.57 $\pm$ 0.03 \\
$H$ & 14.80 $\pm$ 0.03 & -18.18 $\pm$ 0.11 & -18.24 $\pm$ 0.04 \\
$Ks$ & 14.51 $\pm$ 0.05 & -18.45 $\pm$ 0.11 & -18.42 $\pm$ 0.04 \\
\hline
\end{tabular}
\\
$^a$ Apparent magnitude\\
$^b$ Absolute magnitude corrected for reddening\\
$^c$ Average absolute magnitude for SNe~Ia\\  
\label{tab5.1}
\end{table} 

\begin{figure}
\centering
\includegraphics[width=84mm]{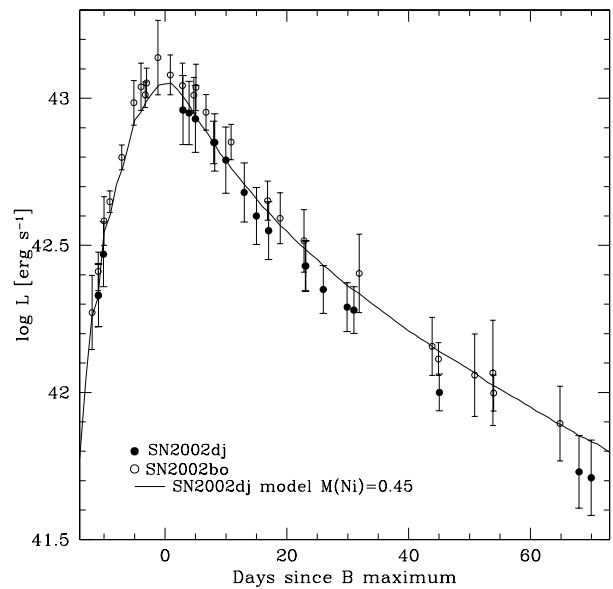}
\caption{Comparison between the $uvoir$ light curves of SN~2002dj
and SN~2002bo. The error bars account only for the uncertainties
in reddening and photometry. The best fit model of SN~2002dj
($^{56}$Ni=0.45$M_{\odot}$) is also shown. }
\label{fig5.1}
\end{figure}

Using the $UBVRIJHKs$ observations of SN~2002dj and adding the UV
contribution determined by  \citet{Suntzeff96}, we constructed a
$uvoir$ light curve that is compared in Fig.~\ref{fig5.1} with that of SN~2002bo and a model generated using a Monte Carlo light curve code
\citep{Cappellaro97}. The curve of SN~2002bo
published by \citet{Benetti04} has been rescaled to the new
distance estimate ($\mu$ = 31.76 $\pm$ 0.07) reported by
\citet{EliasRosa08}. Within the errors the absolute luminosities
of the two SNe are comparable. The model provides a good fit to
the brightness and width near the peak, yielding a $^{56}$Ni mass
of 0.45~$\pm$~0.04 $M_{\odot}$ for SN~2002dj, but after +20 days
it  overestimates the observed flux.  This is probably due to
sudden changes in the opacity at advanced stages which our model
does not properly take into account.
 
\subsection{An intriguing host galaxy}

\begin{table}
\centering
 \caption{Main parameters of SN~2002dj and its host galaxy.}
\hspace{15pt}\\
\begin{tabular}{@{}ll}
\hline
Host galaxy & NGC~5018  \\
Galaxy type & E3 \\ 
RA (2000) & 13$^h$13$^m$01$^s$.7\\
Dec (2000) &$-$19$^{\circ}$31'12''.8 \\
Heliocentric & \\
Recession velocity [km s$^{-1}$] & 2816~$\pm$~1 $\dag$ \\
Distance modulus & 32.93 $\pm$ 0.15 \\
$E(B-V)$ & $0.096 \pm 0.05$  \\
Date of $B$ max (MJD) &  52450.0~$\pm$~0.7\\ 
Offset from the nucleus & $8''.9$ West, $2''.8$ South \\
$\Delta m_{15}(B)$ & 1.08 $\pm$ 0.05\\
stretch factor in $B$ & 0.97 $\pm$ 0.02 \\
\hline 
\end{tabular}
\\
$\dag$ \citep{Rothberg06}
\label{tab5.2}
\end{table}

\begin{table}
\caption{Host galaxies of HVG SNe considered in the paper}
\hspace{15pt}\\
\begin{tabular}{@{}llccl}
 \hline
\multicolumn{1}{c}{SN} & \multicolumn{1}{c}{Galaxy} &  Type &
$\Delta m_{15}(B)$ & \multicolumn{1}{c}{$E(B-V)$} \\
 \hline
SN~1981B &  NGC 4536 & SAB & 1.11 & ~~~~~~$-$\\
SN~1983G &  NGC 4753 & I0 & 1.37 &  ~~~~~~$-$\\
SN~1984A &  NGC 4419 & SB & 1.21 & ~~~~~~$-$ \\
SN~1989A &  NGC 3687 & SAB & 1.06 & ~~~~~~$-$ \\
SN~1997bp &  NGC 4680 & Sp &  1.00 & 0.18(0.04)$^a$ \\
SN~2002bf &  CGCG 266-031 & SB(r)b & $-$ & 0.08(0.04)$^b$ \\
SN~2002bo &  NGC 3190 & Sap & 1.17  &  0.38(0.10)$^c$ \\
SN~2002dj &  NGC~5018 &  E3 & 1.08 & 0.096(0.05)\\
SN~2004dt &  NGC 799 & SB(s) & 1.21 & ~~~~~~$-$ \\
SN~2006X  &  NGC 4321 & SABbc &  1.31 & 1.41(0.04)$^d$ \\
\hline
\end{tabular}  
\\
$^a$ average value between $E(B-V)_{B-V}$ and $E(B-V)_{V-I}$\\
$^b$ From \citet{Leonard05}\\
$^c$  From \citet{Stehle05}\\
$^d$  From \citet{Wang07a}
\label{tab5.3}
\end{table} 

NGC~5018 is the dominant giant elliptical of a small group. It is
peculiar in several respects. First, \citet{Schweizer90}
classified it as one of their best candidates for a recent major
merger. Second, although it is morphologically classified as gE,
its nuclear optical spectrum distinguishes itself by having the
weakest measurement of the absorption line index Mg$_2$ (0.218)
for its velocity dispersion  among over 400 gEs surveyed by
\citet{Davies87}. Its UV (IUE) spectrum lacks the prominent
UV-upturn shortward of 2000\AA, which is typical of old,
metal-rich spheroids \citep{Bertola93}. Nevertheless,  through a
spectral index study, \citet{Leonardi00} found indications for
the presence of a relatively young stellar population ($\sim$2.8
Gyr) in the central region of NGC~5018 with nearly solar
metallicity. A similar age  ($\sim$3Gyr) is proposed by
\citet{Buson04} through a UV  spectroscopy study. \citet{Kim88}
detected a H~I gas bridge connecting NGC~5018 with the nearby
spiral NGC~5022, indicating an ongoing flow toward the giant
elliptical. Furthermore, possible young ($\sim 10^8$
years) globular clusters have been claimed by \citet{Hilker96}.
Finally the H$_{\alpha}$+[N~II] maps reported in
\citet{Goudfrooij94b} reveal the presence of an extended emission
distributed as a strongly warped disk covering the SN~2002dj
position that the authors suggest to be associated with star
forming regions.\\ \citet{Branch93} reported that SNe~Ia
characterized by high expansion velocities, tend to explode in
late type galaxies and suggested that those objects could have
younger progenitors than ``normal'' SNe~Ia.  The HVG SNe analyzed
in this paper confirm this (Table~\ref{tab5.3}). \citet{Hamuy00}
pointed out that slow declining events preferentially occur in
late type galaxies, while fast decliners occur in all type of
galaxies.  A similar result was obtained  by \citet{Sullivan06}
for a sample of high-z SNe~Ia.  These observations seem at odds
with the fact that SN~2002dj was hosted by an elliptical galaxy.
However, given the peculiarities  of NGC5018 it is possible that
SN~2002dj could be associate with a relatively young  stellar
population.

\section{Optical spectra}

\begin{figure}

\centering
\includegraphics[width=84mm]{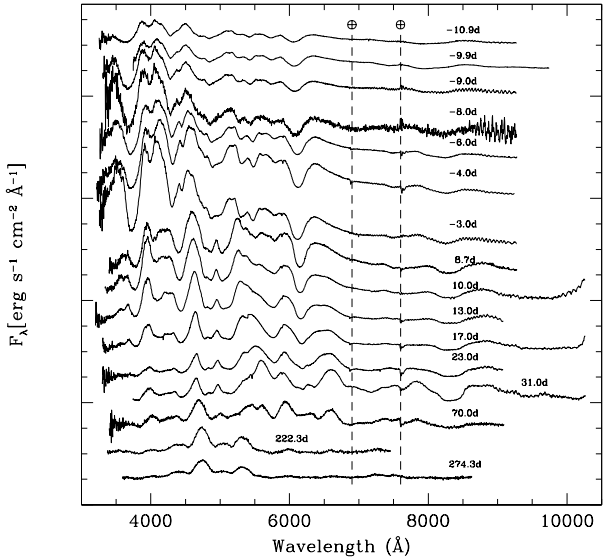}
\caption{Optical spectral evolution of SN~2002dj. With the
exception of the last three spectra, which for presentation
purposes have been multiplied by a constant, the other spectra of
the sequence have been only shifted vertically. The $\earth$ symbol
shows the position of the main telluric features. The spectra are
labeled with the epoch in days past $B$ maximum.}
\label{fig6.1}
\end{figure}

Spectra of SN~2002dj have been obtained at 20 epochs spanning
phases between $-10.9$ and +274 days (Table~\ref{tab6.1}),
following the rapid evolution during the early epochs in detail
and sampling the late phases more sparsely.  The early discovery
of SN~2002dj allows us to compare its spectral evolution
(Fig.~\ref{fig6.1}) with that of SN~2002bo and other SNe,
especially at early epochs.

\begin{table}

\caption{Optical and IR spectroscopic observations of SN~2002dj.}
\begin{tabular}{ccrrrc}

\hline
UT date & M.J.D. & \multicolumn{1}{c}{epoch$^a$} &
\multicolumn{1}{c}{range}  & \multicolumn{1}{c}{res.} &  Instr. \\
     &        & \multicolumn{1}{c}{(days)}    &
     \multicolumn{1}{c}{(\AA)}  & \multicolumn{1}{c}{(\AA)}&         \\
\hline
  02/06/13 & 52439.1   & -10.9  &   3300-9300  &  9 & DFOSC \\
  02/06/14 & 52440.1   &  -9.9  &   3700-9800  & 10 &  EMMI \\
  02/06/15 & 52441.0   &  -9.0  &   3300-9300  &  9 & DFOSC \\ 
  02/06/15 & 52441.1   &  -8.9  &   9400-16500 & 21 &  SofI \\
  02/06/16 & 52442.0   &  -8.0  &   3300-9300  &  9 & DFOSC \\
  02/06/18 & 52444.0   &  -6.0  &   3300-9300  &  9 & DFOSC \\
  02/06/19 & 52444.1   &  -5.9  &   9400-25400 & 21 &  SofI \\
  02/06/20 & 52446.0   &  -4.0  &   3200-9200  &  9 & DFOSC \\
  02/06/21 & 52447.0   &  -3.0  &   3300-9300  &  9 & DFOSC \\
  02/06/24 & 52450.0   &   0.0  &   9400-16500 & 21 &  SofI \\
  02/07/03 & 52458.7   &   8.7  &   3400-9300  &  9 & DFOSC \\
  02/07/04 & 52460.0   &  10.0  &   3300-10300 &  9 & DFOSC \\
  02/07/07 & 52463.0   &  13.0  &   3200-9100  &  9 & DFOSC \\
  02/07/11 & 52467.0   &  17.0  &   3300-10300 &  9 & DFOSC \\
  02/07/11 & 52467.0   &  17.0  &   9400-25200 & 21 &  SofI \\
  02/07/16 & 52473.1   &  23.1  &   3300-9100  &  9 & DFOSC \\
  02/07/25 & 52481.0   &  31.0  &   3700-10300 &  9 & DFOSC \\
  02/09/02 & 52520.0   &  70.0  &   3400-9100  &  9 & DFOSC \\
  03/02/01 & 52672.3   & 222.3  &   3400-7500  & 14 & EFOSC \\
  03/03/25 & 52724.3   & 274.3  &   3600-8600  & 12 & FORS1 \\
\hline
\end{tabular}
$^a$Counted since the time of the $B$ maximum brightness 
M.J.D.=$52450 \pm 0.7$ \\
\label{tab6.1}
\end{table}

\subsection{Premaximum phase}

\begin{figure}
\centering
\includegraphics[width=84mm]{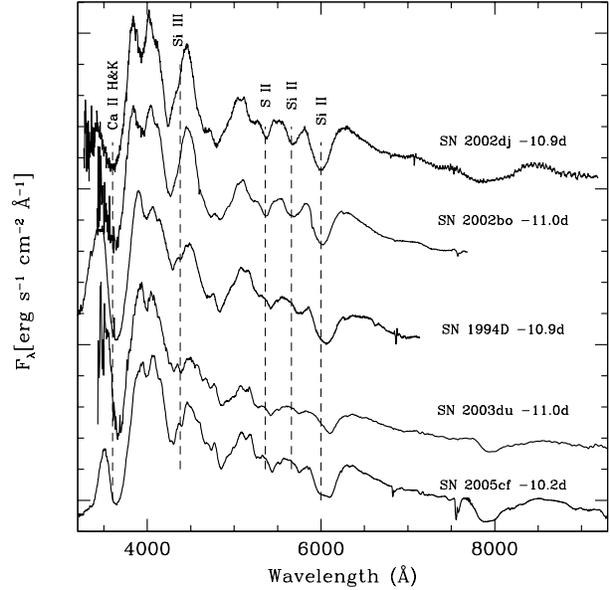}
\caption{Spectrum of SN~2002dj taken at $-$10.9 days. The coeval
spectra of SN~2002bo \citep{Benetti04}, SN1994D \citep{Patat96},
SN~2003du \citep{Stanishev07} and SN~2005cf \citep{Garavini07} are
shown for comparison. The spectra have been corrected for
reddening and redshift.}
\label{fig6.4}
\end{figure}

\begin{figure}
\centering
\includegraphics[width=84mm]{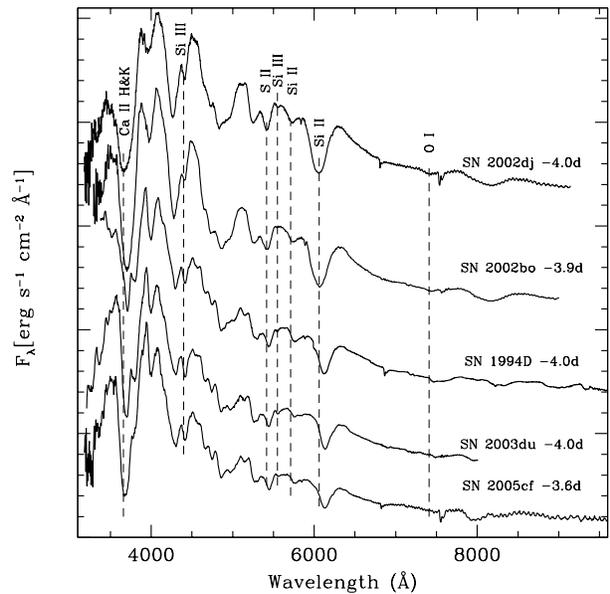}
\caption{Same as Fig.~\ref{fig6.4}, but for a phase 
of $-$4 days. The bibliographic sources for the spectra of SN~2002bo,
SN~1994D, SN~2003du and SN~2005cf are the same as those in
Fig.~\ref{fig6.4}.}
\label{fig6.5}
\end{figure}

In Fig.~\ref{fig6.4} we show the first spectrum of SN~2002dj
obtained about 11 days before $B$ maximum together with coeval
spectra of SN~2002bo, SN~2003du, SN~2005cf and SN~1994D.  Already
at a first glance it is evident how the minima of the Ca~II H\&K,
S~II and Si~II absorption features are more blueshifted in the
spectra of SN~2002dj and SN~2002bo than in LVG SN~2003du,
SN~2005cf and SN~1994D. The Si~II lines are also deeper in HVG SNe.
 The blueshift is due to the higher expansion
velocities of the HVG objects (cf.  Fig.~\ref{fig8.1}).  The
blending of  high velocity detached features \citep{Branch04}
with the photospheric line component could also play an important
role in blueshifting the absorption minimum of Ca~II H\&K and
Si~II $\lambda$6355~\AA~ in HVG SNe as it is for the Ca II IR
triplet (Mazzali et al. 2005, Tanaka et al. 2008). For example,
the minimum of Si~II $\lambda$6355~\AA~ in SN~2002dj and
SN~2002bo is at the same position of the high velocity  component
identified by \citet{Garavini07} in the spectrum of SN~2005cf.
The feature around 4400~\AA~marked in Fig.~\ref{fig6.4} is
attributed to Si~III $\lambda$$\lambda$4553,4568~\AA. The
strength of this line correlates with temperature and is clearly
detected in LVG SNe. The spectrum of SN~2002dj only has a hint
Si~III, and it is completely absent in SN~2002bo. As shown in
Fig.~\ref{fig6.10}, the low temperature of HVG SNe at early
epochs is confirmed by their high $R$(Si~II).  If well traced by
$R$(Si~II) the temperature should rapidly rise in HVG SNe and
quickly reach that of LVG SNe. Indeed at $-$4 days
(Fig.~\ref{fig6.5}) the Si~III is visible with similar strength
in all SNe but SN~2002bo. The latter SN also lacks the weak
absorption at $\sim$5550~\AA~ also attributed to Si~III
($\lambda$5740~\AA) that is present in all other SNe. It is
interesting to note how the minimum of the Si~III line is at
nearly the same wavelength for all SNe, indicating that the line
is formed at roughly the same position in velocity space.  
The Ca~II H\&K doublet (blended with Si~II $\lambda$3858~\AA)
appears to be double-dipped in LVG SN~1994D and SN~2003du, while
it is a single feature in SN~2002dj and SN~2002bo.
\citet{Hatano00} suggested that this difference is related to
higher expansion velocities that lead to a greater degree of line
blending. The Ca~II line blending could also be enhanced by the
presence of high velocity features.  On the other hand
\citet{Lentz00} argued that the double-dip disappears when Si~II
is weaker than Ca~II. Given that in all fast expanding SNe the
Si~II lines are stronger than in the slow expanding SNe, line
blending seems a most consistent explanation.

\subsection{Postmaximum phase}

\begin{figure}
\centering
\includegraphics[width=84mm]{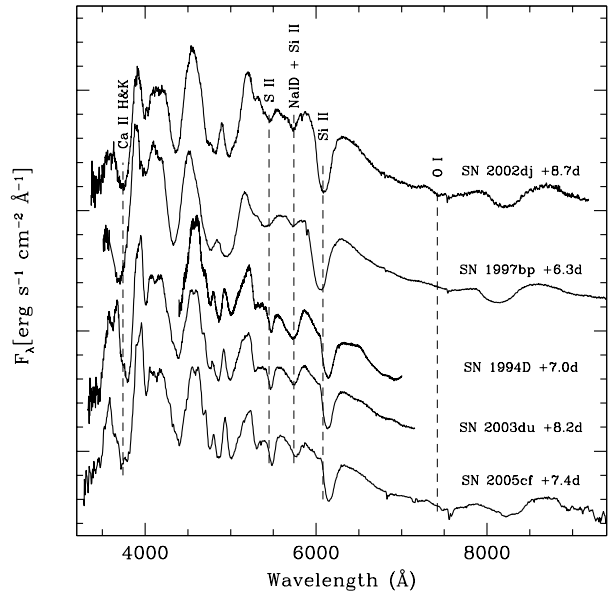}
\caption{Same as Fig.~\ref{fig6.4} for a phase of +9 days. The
bibliographic sources for the spectra are the same as those in
Fig.~\ref{fig6.4} and Fig.~\ref{fig6.5}, while the spectrum of SN~1997bp comes from the
Asiago archive (unpublished).}
\label{fig6.6}
\end{figure}

\begin{figure}
\centering
\includegraphics[width=84mm]{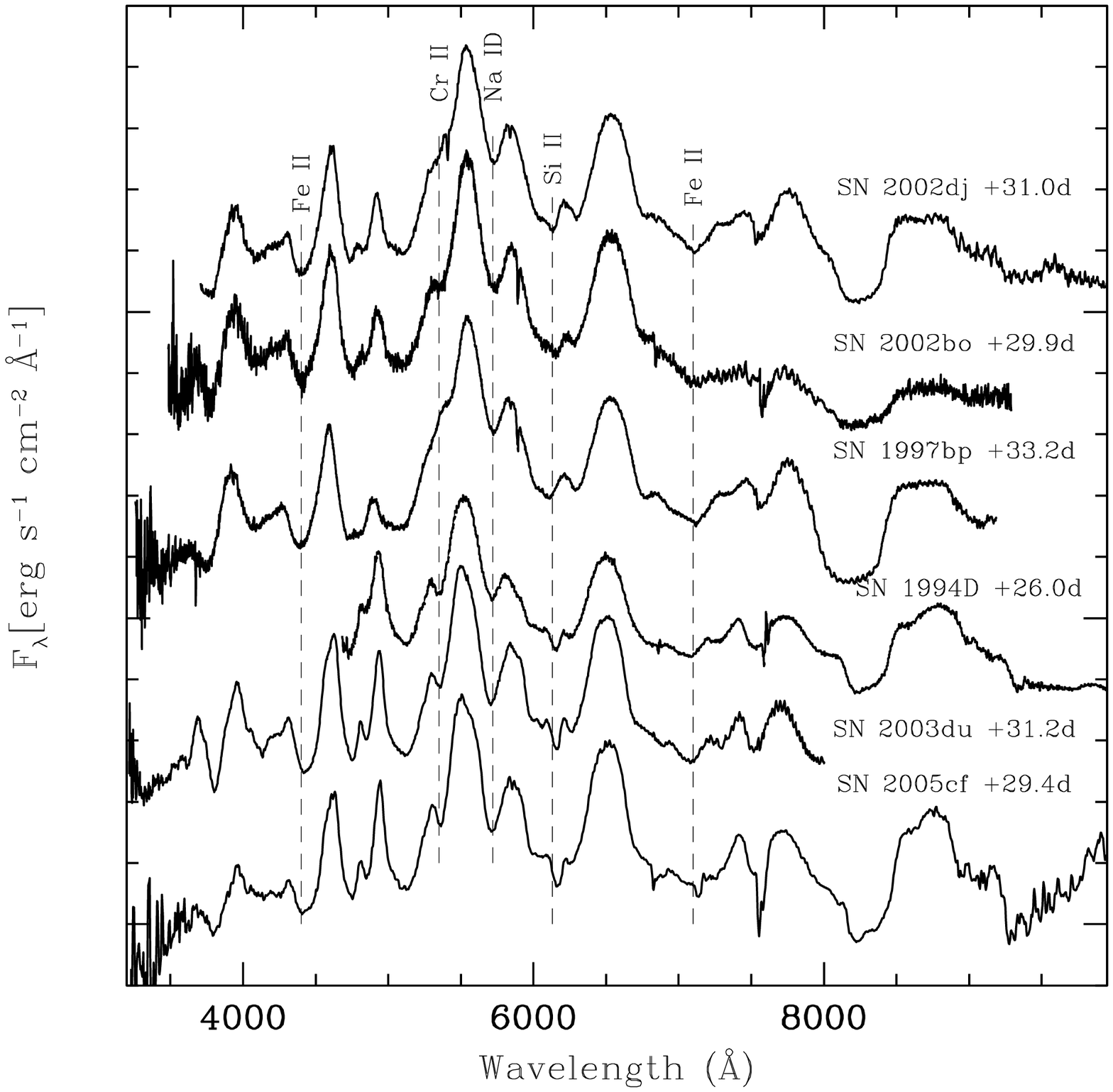}
\caption{Same as Fig.~\ref{fig6.4} for phase +31 days. The
bibliographic sources for the spectra are the same as those in
Fig.~\ref{fig6.5}.}
\label{fig6.7}
\end{figure}

\begin{figure}
\centering
\includegraphics[width=84mm]{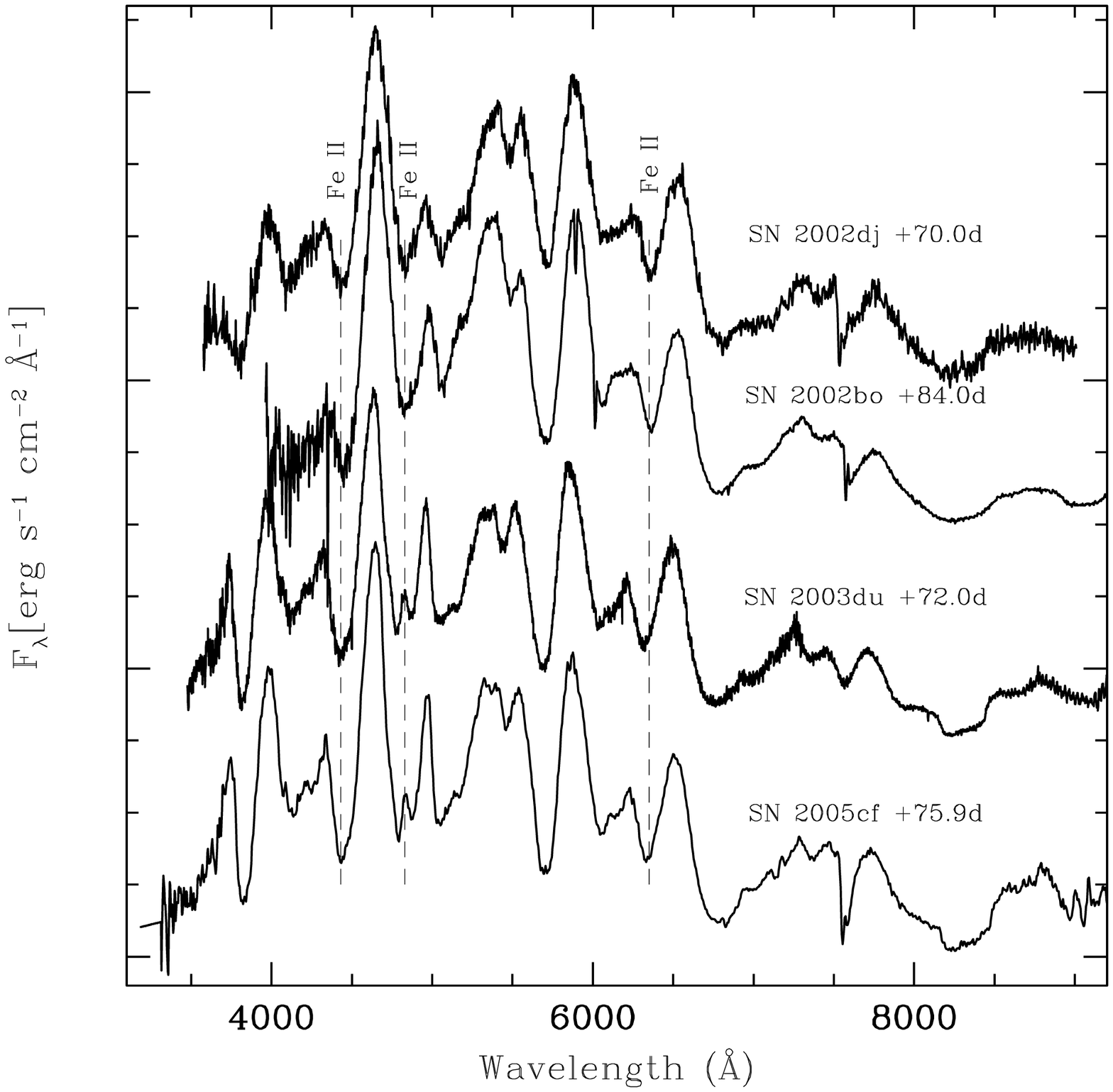}
\caption{Same as Fig.~\ref{fig6.4}, but for phase +70 days. The 
bibliographic sources for the spectra are the same as those in
Fig.~\ref{fig6.4}.}
\label{fig6.8}
\end{figure}

\begin{figure}
\centering
\includegraphics[width=84mm]{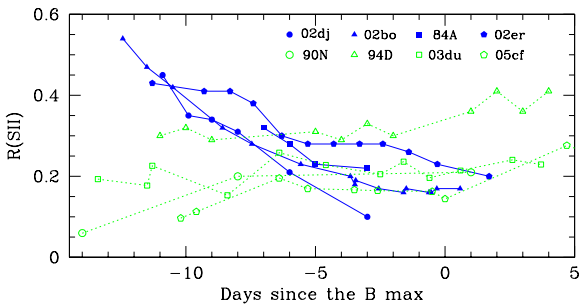}
\caption{Evolution of $R$(SiII) for a sample of HVG SNe (blue
filled symbols) and LVG SNe (empty green symbols).  }
\label{fig6.10}
\end{figure}

A week after maximum the Si~II $\lambda$6355~\AA~ line in the HVG
SN~2002dj and SN~1997bp remains blueshifted with respect to the
LVG SN~1994D, SN~2003du and SN~2005cf (Fig.~\ref{fig6.6}).  The
Si~II $\lambda$6355~\AA~ feature is still deeper in SN~2002dj and
SN~1997bp than in LVG SNe, while, contrary to the pre-maximum
spectra, the S~II line $\lambda$5640~\AA~ is weaker.  SN~2002dj
and SN~1997bp also show less substructure in the region between
4700 and 5100~\AA~ with respect to SN~1994D, SN~2003du and
SN~2005cf.  Finally we note that the absorption minimum of the
feature around 5700~\AA~ at this epoch attributed to a blend of
Na~I~D and Si~II $\lambda$5972~\AA~ \citep{Garavini07} has about
the same position, but in SN~2002dj and especially in SN~1997bp
it is weaker than in the other SNe.  By one month after maximum
the spectra are dominated by iron peak elements
(Fig.~\ref{fig6.7}). The absorption minimum of the lines are
nearly at the same position in all selected SNe. The early fast
expansion of SN~2002dj, SN~2002bo and SN~1997bp is now only reflected
in the broadness of the Ca~II IR triplet. The Na~I~D line is still
weaker than in SN~2003du and SN~2005cf and is now slightly
redshifted. The line at $\sim$5300~\AA~ identified by
\citet{Branch05} as due to Cr~II is barely visible in the faster
expanding SNe while it remains visible in the slower ones.
Finally, around two months after maximum brightness the features
due to Fe~II are slightly redshifted or at the same position in
SN~2002dj and SN~2002bo with respect to SN~2003du and SN~2005cf
(Fig.~\ref{fig6.8}). Also the Na~I~D line is slightly 
redshifted and its intensity is comparable in all SNe.
Interestingly, in the region between 4700 and 5100~\AA~ SN~2002dj
and SN~2002bo still show less substructures. At these epochs line
blending due to the high velocity should not be severe in HVG
SNe, so the lack of lines points to a temperature or abundance
effect.

\subsection{SN~2002dj versus SN~2002bo, a close comparison} 

\begin{figure}

\centering
\includegraphics[width=84mm]{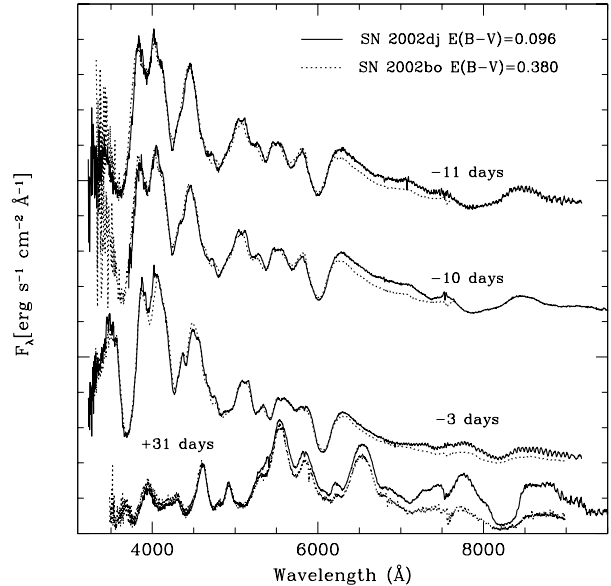}
\caption{Comparison of the reddening and redshift corrected
spectra of SN~2002dj (solid lines) and SN~2002bo (dotted lines).
The epochs labeled in the plot are for SN~2002dj; the SN~2002bo
spectra are coeval within  1.0 day. The SN~2002bo spectra were
scaled in order to  match those of SN~2002dj between 4000 and
6000~\AA.  }
\label{fig6.11}
\end{figure}

We compare the spectra of SN~2002dj with those of SN~2002bo at
four different epochs (Fig.~\ref{fig6.11}). In the wavelength
interval covered by the $B$ and $V$ filters the similarity
between the two objects is evident, but around 6500~\AA~SN~2002bo
shows less flux than SN~2002dj. Since the spectra of both SNe were
calibrated in different epochs with different instruments, we can rule
out an instrumental mis-calibration. Alternatively the difference
could be attributed to an incorrect estimate of the reddening of
the two objects. However, in this case we would expect a gradual
deviation of the two spectra, but the discrepancy looks more like
a break possibly due to a lack of opacity in the photosphere of
SN~2002bo.  For the comparison we chose to obtain the best fit
between 4000 and 6000 \AA. If instead we scale the spectra of
SN~2002bo in order to get the best match with the redder part of
those of SN~2002dj, we end up with a flux excess in SN~2002bo at
blue wavelength. While this is possible, the comparison of the
SN~2002bo colours with those of other LVG and HVG SNe and the
results of the spectral modeling reported by \citet{Stehle05}
point to a flux deficit in the red part instead of an excess in
the blue.

\section{IR spectra}

\begin{figure}

\centering
\includegraphics[width=84mm]{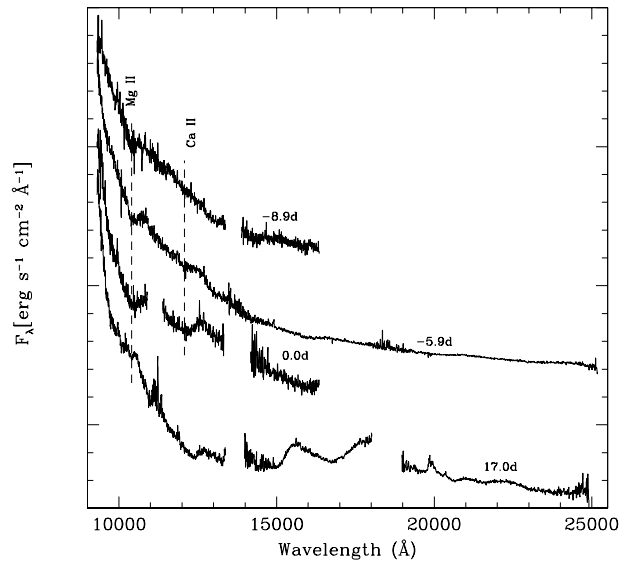}
\caption{Infrared spectral evolution of SN~2002dj. The spectra
are corrected for reddening and redshift.
For graphic exigence the spectra have been multiplied by a suitable factor and vertically
shifted.}
\label{fig7.1}
\end{figure}

Although the sample of published IR spectra of SNe~Ia is growing
rapidly, their number is still small compared with those obtained
at optical wavelengths.  In this context the spectral evolution
of SN~2002dj in the IR from day -8.9 to +17 presented in
Fig.~\ref{fig7.1} constitutes an important contribution to the
global database.

\begin{figure}
\centering
\includegraphics[width=84mm]{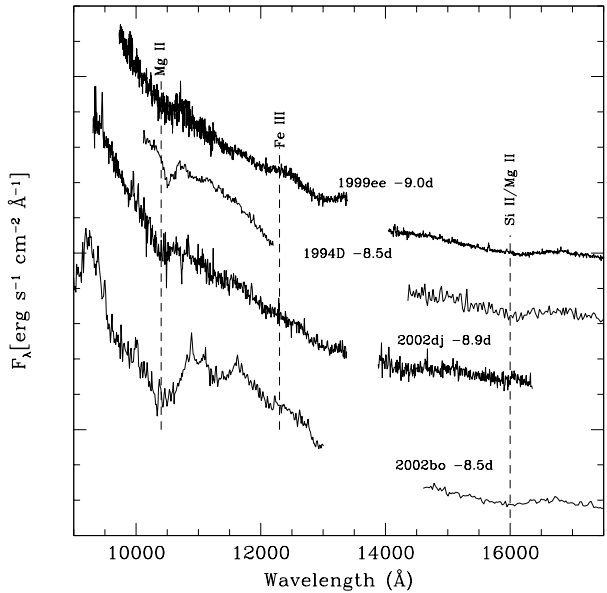}
\caption{IR spectrum of SN~2002dj taken at -8.9 days.  The coeval
spectra of SN~2002bo \citep{Benetti04}, SN1994D \citep{Meikle96},
SN~1999ee \citep{Hamuy02} and SN~2003du \citep{Stanishev07} are
shown for comparison. The spectra have been corrected for
reddening and redshift.}
\label{fig7.2}
\end{figure}

Fig.~\ref{fig7.2} compares the first SN~2002dj IR spectrum with
those of SN~2002bo, SN~1994D and SN~1999ee at similar epochs.
The IR spectrum at this phase is nearly featureless with the
remarkable exception of an absorption at $\sim$10500~\AA. The
identification of this line remains open. It was
attributed by \citet{Meikle96} to He~I $\lambda$10830~\AA~or
Mg~II $\lambda$10926~\AA~ based on an Local Thermodynamic
Equilibrium (LTE) treatment.  Using a NLTE approach,
\citet{Mazzali98a} discussed the conditions under which the He
line could form, and found that its time evolution was at odds
with the observed one.  They therefore favoured either Mg or Si.
A Mg~II identification was also claimed by \citet{Wheeler98}.\\
SN~1999ee shows an emission line at $\sim$12300~\AA\ which
\citet{Rudy02} identified as due to Fe~III in the spectra of
SN~2000cx. The presence of this feature in those SNe is due to
their higher photospheric temperature. As expected, it is absent
in the spectra of both SN~2002dj and SN~2002bo.

\begin{figure}
\centering
\includegraphics[width=84mm]{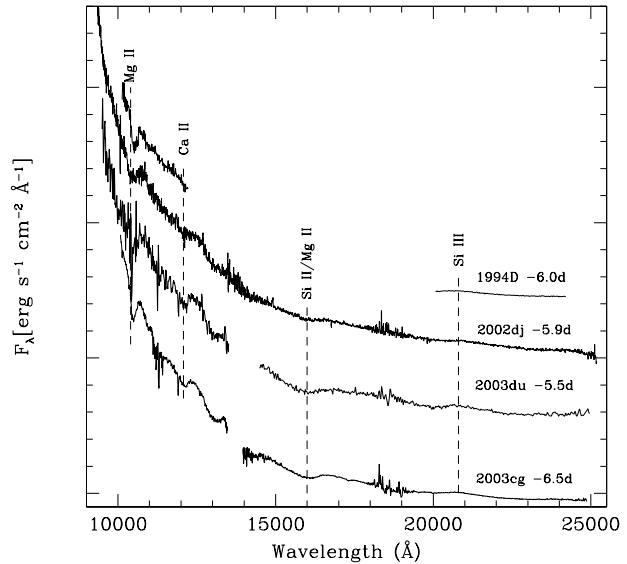}
\caption{Same as Fig.~\ref{fig7.2} for phase -6 days for
SN~2002dj, SN1994D and SN~2003du. The bibliographic sources for
the spectra are the same as those in Fig.~\ref{fig7.2}, while the
spectrum of SN~2003cg was published in \citet{EliasRosa06}.}
\label{fig7.3}
\end{figure}

The  -5.9 day spectrum of SN~2002dj shows a feature around
12100~\AA~ (Fig.~\ref{fig7.3}) which was not visible in the earliest spectrum (cf.
Fig.~\ref{fig7.1}).  \citet{Marion03} identified this feature as
a blend of several Ca~II lines between 12430~\AA~and
12830~\AA. Like for the Mg~II line, the blueshift of the
Ca~II minimum is larger in SN~2002dj than in all other SNe.
Finally at $\sim$20800~\AA\ a weak emission is visible in the
spectra of SN~2002dj, SN~2003du and SN~2003cg attributed to
Si~III by \citet{Wheeler98}.

Following \citet{Rudy02} we fitted the two earliest IR spectra of
SN~2002dj to estimate the black body temperature.  We used only
sections of the spectrum free of telluric features or absorption
lines.  Assuming $E(B-V)$=0.096 for SN~2002dj we obtain T=11900~K
at -8.5 days compared with  T=12900~K and T=23500~K for SN~1994D
and SN~1999ee reported in Fig.~\ref{fig7.2}.  For the latter two
SNe we assumed $E(B-V)$=0.06~$\pm$~0.02 \citep{Patat96} and
$E(B-V)$=0.30~$\pm$~0.04 \citep{Stritzinger02}, respectively. The
high temperature of SN~1999ee ($\Delta m_{15}(B)=0.96 \pm 0.02$)
is consistent with the presence of the Fe~III emission feature in
its spectrum. It is also similar to the 25000~K temperature of
SN~2000cx \citep{Rudy02} taken at -8.0 days, which shows a
stronger Fe~III emission than SN~1999ee.  This gives us
confidence that the technique provides reliable results, when
applied to early time spectra. The sequence of increasing
temperature from SN~2002dj to SN~1999ee confirms the results
obtained from $R$(Si~II) (Fig.~\ref{fig6.10}). Indeed, for the
SN~2002dj at -5.9 days we found a much higher temperature
(T=20500~K). This confirms its fast rise as traced by the
$R$(Si~II) and the appearance of a Si~III line in the $K$ band in
the spectrum of SN~2002dj.

\begin{figure}
\centering
\includegraphics[width=84mm]{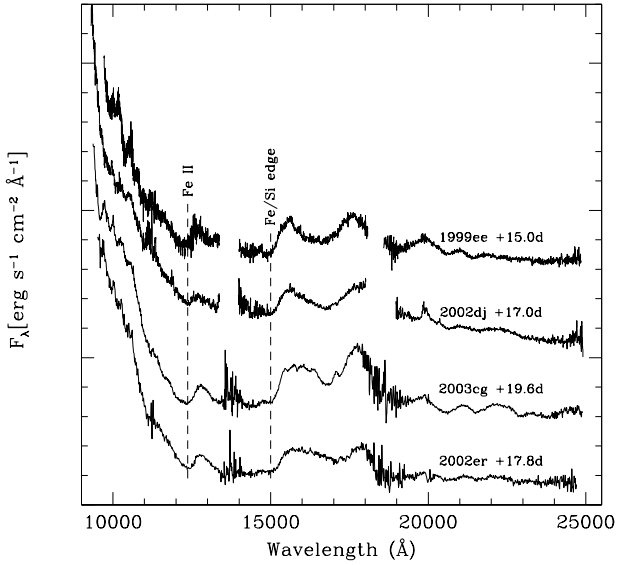}
\caption{Same as Fig.~\ref{fig7.2} for phase +17.0 days. The
bibliographic sources for the spectra are the same as those in
Fig.~\ref{fig7.2} and Fig.~\ref{fig7.3}. The spectrum of
SN~2002er that was published by \citet{Kotak05}.}
\label{fig7.4}
\end{figure}

\citet{Rudy02} proposed that the black body IR spectrum fit can
be used to derive also an estimate of absorption. We note,
however, that 
the resulting $\chi^2$ surface for temperature and colour excess
is nearly flat providing only weak constraints. At +17 days we
see that several lines populate the IR spectrum of SN~2002dj and
those of other SNe (Fig.~\ref{fig7.4}). In particular, a broad
P-Cygni feature attributed to Fe~II \citep{Marion03} is observed
at $\sim$12350~\AA. In SN~2002dj the blueshift of its minimum is
roughly the same as in SN~2003cg and SN~2002er, while SN~1999ee
displays higher velocities. Finally, \citet{Marion03} suggested
that the minimum around $\sim$15000~\AA~defines the transition
from partial to complete silicon burning. The fact that the
velocities inferred by the position of this minimum is the same
in SN~2002dj, SN~2003cg and SN~2002er indicate that these SNe
have an iron core of similar size.

\section{Expansion velocities}

\begin{figure}

\centering
\includegraphics[width=84mm]{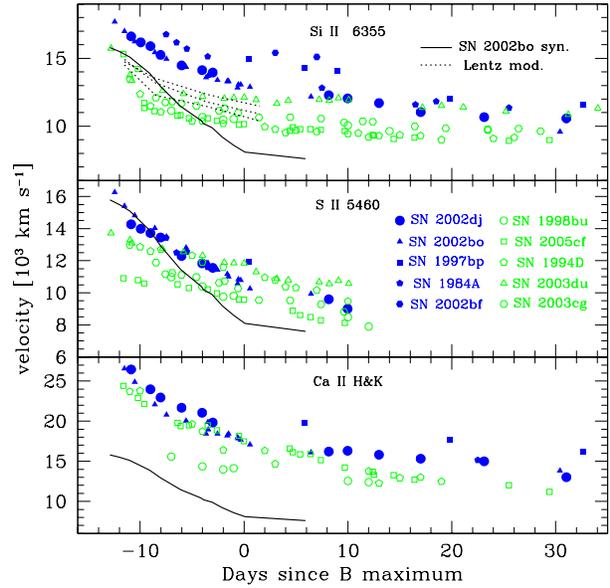}
\caption{Velocity evolution of Si~II (top panel), S~II (middle panel)
and Ca~II (bottom panel) in SN~2002dj, SN~2002bo,
SN~1997bp, SN~1984A, SN~2002bf, SN~1998bu, SN~2003du, SN~2005cf,
SN~1994D and SN~2003cg. The solid lines represent the evolution
of the photospheric velocity of SN~2002bo computed by
\citet{Stehle05}, while the dashed lines show the velocities
predicted by \citet{Lentz00} model for the case of $\times$10 (top
line at epoch 0), $\times$3 (middle line) and $\times$1/3 (bottom
line) C+O solar metallicity.}
\label{fig8.1}
\end{figure}

In Fig.~\ref{fig8.1} we show the expansion velocities measured
from the blueshift of the minima of the Si~II
($\lambda$6355~\AA), S~II ($\lambda$5460~\AA) and Ca~II (H\&K).
It is evident that at early epochs the HVG SN~1984A, SN~1997bp,
SN~2002bo and SN~2002dj are faster than LVG SNe in all those
lines. The two groups are separated most strongly in Si~II
velocity.  Nevertheless, as noted by \citet{Jeffery90} strong
features like Si~II ($\lambda$6355~\AA) reliably trace the
photospheric velocity only at early epochs, when the mass above
the photosphere is small. For HVG SNe like SN~2002dj and
SN~2002bo, where Si~II is particularly strong and where high
velocity features could play an important role \citep{Mazzali05},
this is unlikely even at very early times. In SN~2002bo for
example the photospheric evolution computed by \citet{Stehle05}
even at -13 days has a velocity $\sim$2000 km s$^{-1}$ lower than
that measured from the minimum of the Si~II line (cf.
Fig.~\ref{fig8.1} upper panel). The relatively weak S~II
(5460~\AA) seems to be a better tracer of the photospheric
velocity. In HVG SNe this feature is at higher velocities than in
LVG SNe, but due to its steep velocity decline it approaches the
LVG group already before maximum.  If we take S~II as a proxy for
the photospheric velocity, its evolution  may indicate that at
very early epochs the photosphere in HVG SNe is located at larger
radii than in LVG SNe, but moving inward more quickly, it reaches
a comparable position already before maximum. For Ca~II (H\&K),
some LVG SNe closely resemble the velocity evolution of HVG SNe,
making this feature unsuitable to separate the two groups.

\begin{figure}

\centering
\includegraphics[width=84mm]{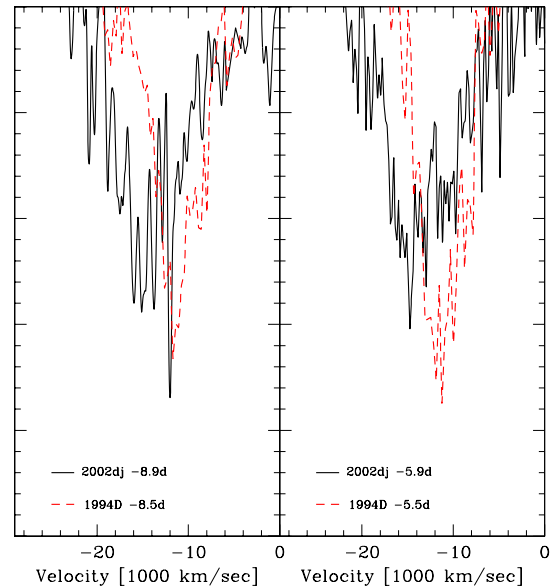}
\caption{Comparison between the Mg~II $\lambda$10926 feature of
SN~2002dj (solid line) and SN~1994D (dashed line) at $\sim -$8.5
days (left panel) and -$6.0$ days (right).} \label{fig8.2}
\end{figure}

In the case of Mg~II (10926~\AA), the expansion velocity measured
from the absorption minimum at -8.5 days is $\sim$15000 km
s$^{-1}$ for SN~2002dj and SN~2002bo and significantly slower
($\sim$11000 km s$^{-1}$) for SN~1994D.  The feature appears
deeper in SN~1994D than in SN~2002dj (Fig.~\ref{fig8.2}).
We do not observe a change in the line velocity from the first to the
 second epoch in both SNe and infer that even earlier than one week before maximum  the photosphere is below the magnesium layer.  Since Mg~II is produced
 in the outer layers by oxygen burning \citep{Wheeler98}, this means
 that only IR spectra observed earlier than this may tell us about the
 boundary between explosive carbon and oxygen burning.
Nevertheless, we note in Fig.~\ref{fig8.2} that the blue wing of
the Mg~II line extends up to $\sim$19000 km s$^{-1}$ in SN~2002dj
and only up to $\sim$15000 km s$^{-1}$ in SN~1994D. Hence, the
burning front must have propagated further out in SN~2002dj than in
SN~1994D.

\section{Spectral Modelling}

\subsection{Early phase}

\begin{figure}
\centering
\includegraphics[width=60mm,angle=270]{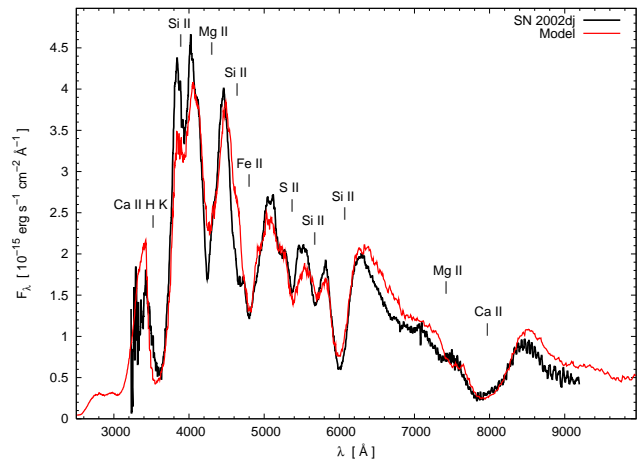}
\caption{Comparison between the first spectrum of SN~2002dj taken
at -10.9 days and a synthetic spectrum. The the input parameters
for the latter are $E(B-V)$=0.096, v$_{ph}$=11600 km s$^{-1}$,
log(L)=42.33.}
\label{fig9.1}
\end{figure}

We modeled the earliest spectrum of SN~2002dj  obtained 10.9 days
before B maximum using the Montecarlo code described in Mazzali
\& Lucy (1993), Lucy (1999) and Mazzali (2000), modified to
include abundance stratification as described in
\citet{Stehle05}.  While a good fit to the overall spectrum can
be obtained for the input parameters reported in the caption of
Fig.~\ref{fig9.1}, it is not possible with the density structure
of W7 \citep{Nomoto84} to reproduce the high velocity features.
One modification that we adopted here to overcome this is to
increase the mass at high velocity. In particular, we assume the
presence of 0.04$M_{\sun}$ of material at velocities above 18000
km s$^{-1}$. This may be caused by interaction of the outer
ejecta with circumstellar material. \citet{Tanaka08} discuss this
and other possibilities in detail. Let us note that the model
suffers only to a minor degree from the problem of overfitting
the flux to the red of the Si~II ($\lambda$6355~\AA) line
suggesting high line opacity in all regions. This can not be the
case for SN~2002bo where the observed spectrum is noticeably
below the model \citep{Stehle05}.

Finally, we note that the spectral modelling seems to exclude the
presence of C~II in the earliest spectrum of SN~2002dj. The
presence of this high velocity feature was found in the -9 days
spectrum of SN~1998aq \citep{Branch03} and in the spectra of
SN~2006gz from -14 to -10 days \citep{Hicken07}. High velocity
C~II ($\lambda$6578~\AA) in the -14 days spectrum of SN~1990N was
also claimed by \citet{Fisher97}, although \citet{Mazzali01}
showed that a high-velocity component of Si~II
($\lambda$6355~\AA) was another viable identification. Moreover,
the presence of small quantities of carbon 
has been shown to improve the fit of the very early
spectra of SN~1994D \citep{Branch05}, SN~2001el and SN~2003du
\citep{Tanaka08}.  The latter SN~Ia all are LVG SNe, therefore
the absence of C~II in the -10.9 days spectrum of SN~2002dj and
in the -12.9 days spectrum of SN~2002bo \citep{Stehle05} could
indicate that the carbon burning has penetrated further out in
these explosions \citep{Stehle05}.

\subsection{Nebular phase}

\begin{figure}

\centering
\includegraphics[width=84mm]{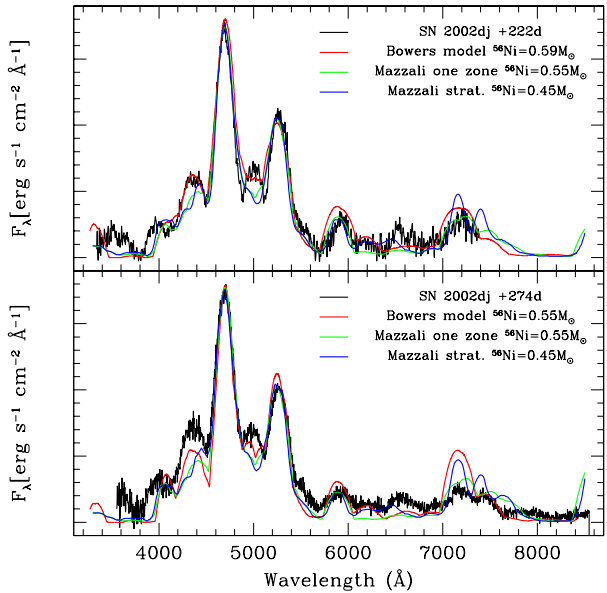}
\caption{Nebular spectra of SN~2002dj taken at +222 days (top
panel) and +274 (bottom panel) compared with various synthetic
spectra. }
\label{fig9.2}

\end{figure}

In order to estimate the errors on the $^{56}$Ni mass estimate
obtained by modelling nebular spectra, we generated
a set of synthetic spectra computed using two different codes and
two different assumptions of the $^{56}$Ni mass distribution. The six synthetic spectra together with the two observed spectra of SN 2002dj  are presented in Fig.~\ref{fig9.2}. Both codes assume line-formation in non-local thermodynamic
equilibrium (non-LTE) conditions.  The code described in
\citet{Bowers97} assumes a uniform density, homologously expanding
sphere containing iron, cobalt and sulphur, with relative
abundances modulated by the radioactive decay.  In computing the
SED only singly and doubly-ionized species are considered because
they are predicted to dominate the spectrum.  The code described
in \citet{Mazzali01} can use either a homologously expanding
nebula of finite extent, uniform density and composition, as in
\citet{Bowers97}, or stratification in density and
abundance \citep{Mazzali07}.  In this case, the density profile
is taken from the explosion model W7 \citep{Nomoto84}.
$\gamma$-rays and positrons are emitted at various depths
according to the distribution of $^{56}$Ni and their propagation
and deposition is followed using a Monte Carlo scheme similar to
that discussed by \citet{Cappellaro97} for their light-curve
models.  Gas heating and cooling as well as line emissivity are
computed in non-LTE in each radial shell, and the line
profiles are computed assuming each shell to contribute to a
truncated parabola, with an inner truncation point corresponding
in velocity to the inner boundary of the shell considered. The
sum of these truncated parabolas is the emerging spectrum.
Constant $\gamma$-ray ($\kappa_{\gamma} =
0.027$\,cm$^2$\,g$^{-1}$) and positron opacities ($\kappa_{e^+} =
7$\,cm$^2$\,g$^{-1}$) are assumed in both cases.

The synthetic spectra yield slightly different results for the
$^{56}$Ni mass, reflecting the different spectral calibrations
and code assumptions. The \citet{Bowers97} model gives a
$^{56}$Ni mass of 0.59~M$_{\sun}$ and 0.55~M$_{\sun}$ for the
+222d and +274d spectra, respectively, while the one-zone model
from \citet{Mazzali01} yields 0.55~M$_{\sun}$ for both epochs.
With the \citet{Mazzali07} stratified model we obtain
0.45~M$_{\sun}$ for both spectra.  Results obtained with the same
code on the two different epochs are consistent, indicating the
robustness of the fit. On the other hand, both one-zone models
yield a $^{56}$Ni mass slightly larger than the stratified model.
\citet{Ruiz-Lapuente95} found similar results  with their stratified nebular model as opposed to the one-zone model
\citep{Ruiz-Lapuente92}.
The main difference between the models is that in the one-zone
codes the $^{56}$Ni density is uniform, while in the stratified
code it is higher near the centre, which leads to a higher
$\gamma$-ray deposition at late times and reduces the $^{56}$Ni
mass necessary to produce a given luminosity. The $^{56}$Ni mass
obtained with the stratified model is in excellent agreement with
the one obtained from the $uvoir$ light curve, suggesting that
this assumption provides a good description of the physical
conditions in the nebula.

\section{Discussion and conclusion}

We have presented optical and IR photometric and spectroscopic
observations of the HVG Type Ia SN~2002dj. The ejecta kinematics
of this object are very similar to that of SN~2002bo. The direct
comparison of the line shapes and strengths of the two objects
reveals an even more impressive similarity. The one obvious
distinction is a peculiar flux deficit redward of 6500~\AA~ in SN~2002bo.

The light curves are nearly identical with the notable exception
of the $B$ band at epochs later than +40 days. However, the
$V-IR$ colours are remarkably different. SN~2002bo is much bluer
confirming the lack of flux redward of $\sim$6500~\AA.  Finally,
also the bolometric absolute luminosity around maximum of the two
objects turns out to be very similar. 
The IR contribution is in in fact very small at those epochs
\citep{Suntzeff96}.

Starting from
the photometric similarities of SN~2002dj and SN~2002bo, we
studied the characteristics of other HVG SNe.
Although
the number of objects for which reliable multi-band
photometry covering both very early and late light curves phases
is limited, the HVG SNe analyzed in this paper seem to share
the following photometric characteristics if
compared with LVG SNe:

\begin{enumerate}
\item[1)] A general fast rise to maximum brightness in all filters
\item[2)] More pronounced inflections in the $V$ and $R$ bands around
+25 days 
\item[3)] A brighter, more slowly declining $B$ light curve after 
+40 days
\item[4)] Possibly different colours 
\end{enumerate}

To complete the picture these properties should be combined with
parameters one can derive from the spectral evolution:


\begin{enumerate}
\item[1)] Fast expanding ejecta
\item[2)] Strong absorption due to intermediate mass elements (IME) at early phases
\item[3)] Absence of carbon in the very early spectra
\item[4)] Early time low photospheric temperature rapidly
increasing toward maximum brightness
\item[5)] Nebular spectra with kinematics not so different or
even slower than that of LVG SNe.  
\end{enumerate}

\citet{Benetti04} suggested that the high expansion velocities of
IME are due to the fact that in HVG SNe the burning front extends
further out into the outermost layers than in LVG SNe.  This
hypothesis is supported by the extension of the blue wing of the
Mg~II line at velocities close to 19000 km sec$^{-1}$ observable
in the early IR spectra of SN~2002dj and by the absence of carbon
in the -10.9 spectrum of SN~2002dj and -12.9 spectrum of SN~2002bo. In this
scenario the density of IME is enhanced in the layers where
usually unburned carbon and oxygen are present.  The IME density
enhancement in the outermost layers would induce stronger
absorptions at early epochs and increase the opacity in these
regions, moving the photosphere to larger radii. The blending
between high velocity features and the photospheric component
could also contribute to the absorption enhancement
\citep{Tanaka08}. Since the photosphere is located at high
velocity shells, this could justify both the fast rise to maximum
observed in the light curve and, because the heating produced by
the $\gamma-$rays is less effective, the lower
temperature in the line forming region.  Due to their high
velocity the density in those shells decreases rapidly and hence
the photosphere quickly recedes into the ejecta. Indeed, around
maximum, the light curves of SN~2002dj, SN~2002bo and SN~2006X
approached those of LVG SNe. Also the temperature is nearly
identical as traced by the $R$(Si~II) evolution and by the
appearance of the Si~III lines, the absorption minimum of which
is at the same position in all SNe. The initially low and then
rapidly increasing temperature of SN~2002dj was confirmed by
fitting a black body to its two earliest IR spectra. 

We note that in HVG SNe more pronounced inflections in the $V$
and $R$ light curves around +25 days seem to correlate with
expansion velocities. The fastest expanding SN~1997bp, SN~2002bf
and SN~2006X display a stronger inflection than in SN~2002bo and
SN~2002dj, where the effect is barely visible.  We must stress
that in the case of SN~2006X, due to the large reddening, the
intensity of this feature has to be taken with caution.  Also the
brighter $B$ light curve after +40 days seems to correlate with
the expansion velocity. \citet{Wang07b} suggested that SN~2006X
declined more slowly after +40 days and could be due to a light
echo produced by the circumstellar dust.  In the case of
SN~2002dj there is no clear evidence of dust obscuring its light. 
but it is also the HVG SNe with the $B$ light curve closer to ``normal'' SNe Ia.
Nevertheless, SN~1997bp and especially SN~2002bf with low
absorption closely resemble the behavior of SN~2006X having also
comparable expansion velocities.  Moreover, part of the reddening
estimated for those SNe could be biased due to their peculiar
colour and colours evolution.

Regarding the colours of HVG SNe we note that \citet{Dominguez01}
using a set of Delayed Detonation models (DD) have found that
increasing the metallicity of the progenitor white dwarf, the
$B-V$ colour of the resulting SN~Ia becomes redder.
Unfortunately, this study does not extend the results to other
bands and is limited to a metallicity Z=0.02. However,
\citet{Timmes03} have proposed that the effect of the progenitor
metallicity is significantly enhanced at solar metallicity and
above. In particular Fig.~9 of Dominguez et al.  (2001) shows
that passing from Z=0.001 to Z=0.02 the flux in $B$ is depressed,
while that in $V$ is enhanced.  If this is maintained at higher
metallicity it could be the source of the red $B-V$ and blue
$V-I$ colours we found in SN~2002bo, SN~1997bp and especially
SN~2002dj where the effect of reddening should be negligible.
Furthermore, Timmes et al. (2003) have pointed out that in metal
rich progenitors due to the more efficient electron capture, more
$^{54}$Fe and $^{58}$Ni are produced at the expense of $^{56}$Ni.
\citet{Mazzali06} used this finding to explore the effect on
spectra and light curves by varying the fraction of $^{54}$Fe and
$^{58}$Ni produced in burning to nuclear statistical equilibrium
and concluded that the colours of SN~Ia with a larger fraction of
these elements (i.e.  the more metal rich ones) have redder $B-V$
colour.  Moreover, indications of larger $^{54}$Fe and $^{58}$Ni
and lower $^{56}$Ni production for the HVG SN~1981B, SN~2002bo
and SN~2002dj with respect the similar $\Delta m_{15}(B)$ LVG
SN~1990N, SN~2003du and SN~2003cg have been found by
\citet{Mazzali07}. 
\citet{Lentz00} pointed out that increasing the abundance of all
elements heavier than oxygen in the unburned C+O layer of the
pure deflagration model W7 \citep{Nomoto84} leads to a larger
blueshift of SN~Ia absorption lines.  This is due to the fact
that the higher opacity provided by the large amount of metals
confines the line forming region to the outer (faster) part of
the ejecta. Although, this supports the hypothesis of metal rich
progenitors for HVG SNe, we favour the configuration with an
extended burning front to explain the HVG SNe observables.

To further complicate the picture we must mention that 
\citet{Hoeflich98} found that SN~Ia with metal rich progenitors
should have a slower rise time to maximum than metal poor ones
clearly contrary to what we found of SN~2002dj, SN~2002bo and
SN~2006X. 
Given the very small sample of HVG SNe, the uncertainties
afflicting their colours and the model shortcomings we briefly
recapped here, our suggestion of metal rich progenitor for HVG
SNe are tentative and provide possible avenues to explore 
when more data on HVG SNe and more sophisticated models will be
available.

Fast expanding SNe are rare both at low and high redshift
\citep{Blondin06, Balland07, Hook05}. The systematic error they
could introduce into cosmological studies due to their peculiar
colours appears to be negligible.  Detailed analysis will become
possible when a larger sample of well-observed rapidly evolving
SNe Ia will be available.

\section*{Acknowledgments}
G.P acknowledges support by the Proyecto FONDECYT 3070034. This
work was also supported by the European Community's Human
Potential Programme under contract HPRN-CT-2002-00303, ``The
Physics of Type Ia Supernovae''. It is partially based on
observations made with ESO Telescopes under programme ID
169.D-0670. This work is also based on observations performed at
the Jacobus Kapteyn Telescope (JKT) and the Isaac Newton
Telescope of the Isaac Newton Group at La Palma, Spain, the
Nordic Optical Telescope at La Palma, Spain, and the 0.9m, 1.0m
telescopes at Cerro Tololo Interamerican Observatory.  A.C., M.H.
and J.M. acknowledge support from Centro de Astrofisica FONDAP
15010003, and N\'ucleo Milenio P06-045-F funded by Programa
Bicentenario de Ciencia y Tecnolog\'ia from CONICYT and Programa
Iniciativa Cient\'ifica Milenio from MIDEPLAN.  A.C. also
acknowledges support from Proyecto FONDECYT 1051061.  M.H.
further acknowledges support from Proyecto FONDECYT 1060808.  SB,
EC and MT are supported by the Italian Ministry of Education via
the PRIN 2006 n.2006022731 002.  This work has made use of the
NASA/IPAC Extragalatic database (NED) which is operated by the
JET Propulsion Laboratory, California Institute of Technology,
under contract with the US National Aeronautic and Space
Administration.

\appendix

\section{S-correction}
To avoid the well known limitations of the colour equation to
calibrate SN, we used the following calibration path to transform
the SN instrumental magnitude into a standard photometric system.
To simplify the explanation of the various steps, we will discuss
as an example the calibration of the $B$ band and we will use
only one of the local standard stars. First one needs to
transform the standard magnitude of the sequence star in the
natural photometric system of the instrument using the following
formula:

\begin{equation}
B_{nat}^*=B^*-\gamma^B_{B-V}(B^*-V^*)
\label{equa1} 
\end{equation}

\noindent
where $B^*$ and $V^*$ are the standard magnitude of the local
standard star,
while $\gamma^B_{B-V}$ is the colour term.  Then one computes
the instrument zero point in the natural photometric system
$ZP(B)_{nat}$=$B_{nat}^*-b^*$, where $b^*$ is the local standard star
instrumental magnitude.

Finally we calibrated the magnitude of the SN using the equation:

\begin{equation}
 B^{SN}=ZP(B)_{nat}+b^{SN}+\delta_B
\label{equa2} 
\end{equation}

\noindent
where $b^{SN}$ is the instrumental magnitude of the SN. We did not
correct for atmospheric absorption because we assume that the SN
and local standard stars are at the same airmass and hence the
absorption is already included in $ZP(B)_{nat}$. We define $\delta_B$
as the difference between the synthetic magnitude of the SN computed
using the standard and the natural system passbands.

\begin{equation}
\delta_B=B_{sy}^{SN}-b_{sy}^{SN}
\label{equa3} 
\end{equation}

\noindent
with:

\begin{equation}
B_{sy}^{SN}=-2.5log\frac{\int_{-\infty}^{+\infty} N(\lambda)S_{st}^B(\lambda)d \lambda}{\int_{-\infty}^{+\infty}S_{st}^B(\lambda)d \lambda} + ZP(B)_{sy}
\label{equa4} 
\end{equation}

\begin{equation}
 b_{sy}^{SN}=-2.5log\frac{\int_{-\infty}^{+\infty} N(\lambda)S_{nat}^B(\lambda)d \lambda}{\int_{-\infty}^{+\infty}S_{nat}^B(\lambda)d \lambda} + ZP(b)_{sy}
\label{equa5} 
\end{equation}

where $N(\lambda)$ is the photon number distribution of the SN,
$S_{nat}^B(\lambda)$ and $S_{st}^B(\lambda)$ are the instrumental
and standard $B$ passbands.  In the equations \ref{equa4} and
\ref{equa5}, $N(\lambda)$ is used instead of $F(\lambda)$ because
CCD and IR detector are photon counting devices.  The
\citet{Bessell90} passbands were also reported in photon unit
dividing the original ones for the wavelength, while the
\citet{Persson98} bands are already in photon units. Since in the
spectra which are used to compute the synthetic magnitudes the
flux is expressed in energy units, $N(\lambda)$ becomes:

\begin{equation}
N(\lambda)=\frac{F(\lambda)\lambda}{hc}
\label{equa9} 
\end{equation} 

The zero points $ZP(B)_{sy}$ and $ZP(b)_{sy}$ were computed using
a subset of spectrophotometric standard stars for which
accurate photometry is available. In particular, for
$ZP(B)_{sy}$ the standard magnitude was used directly, while for
$ZP(b)_{sy}$ the magnitudes were previously reported in the
instrument natural photometric system using equation
(\ref{equa1}).  Putting together equations \ref{equa1},
\ref{equa2} and \ref{equa3} we obtain:

\begin{equation}
 B^{SN}=B^*+b^{SN}-b^*-\gamma^B_{B-V}(B^*-V^*)+B_{sy}^{SN}-b_{sy}^{SN}
\label{equa8} 
\end{equation} 


\begin{figure}
\centering
\includegraphics[width=84mm]{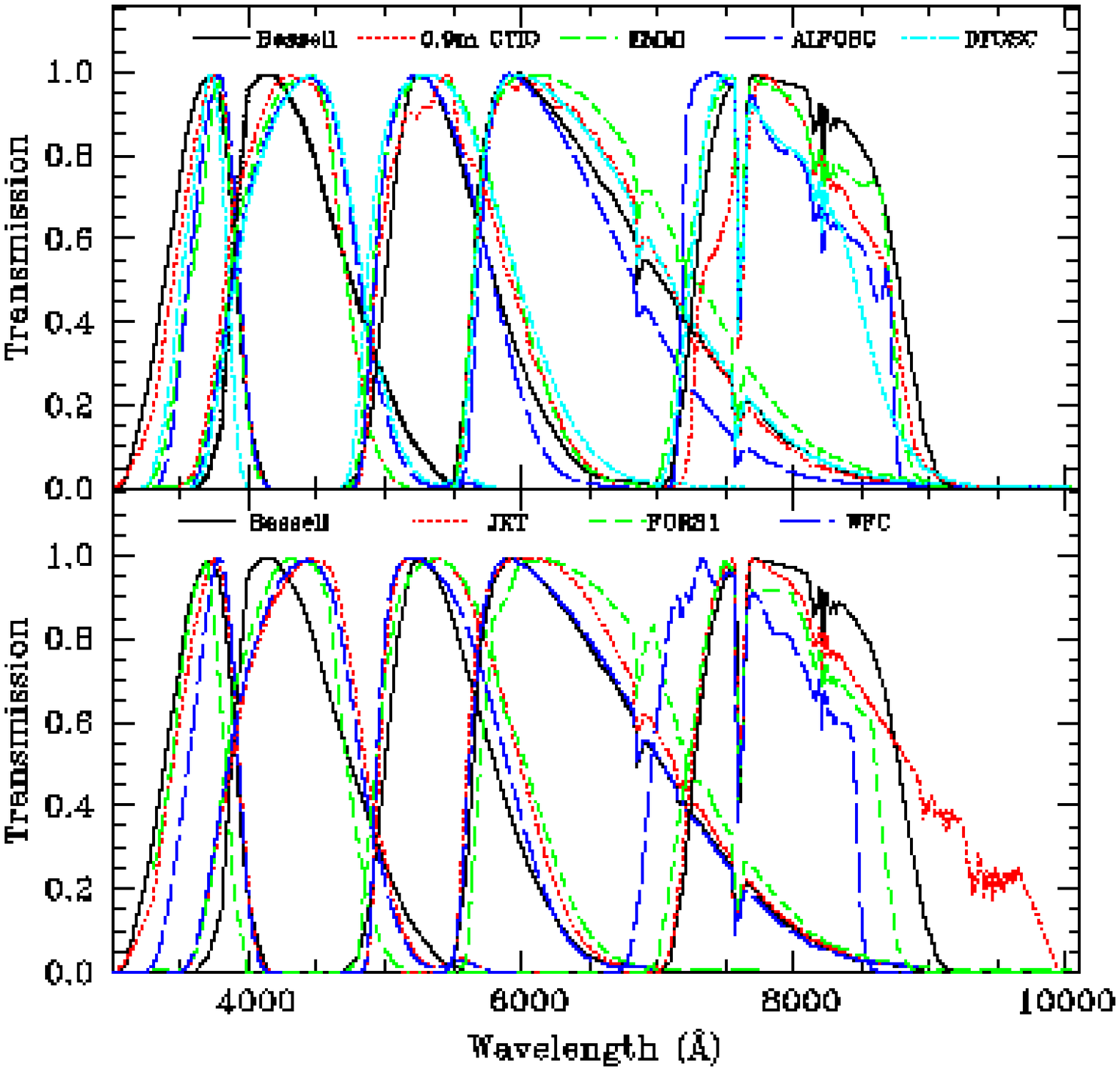}
\caption{Comparison of the different instrumental $UBVRI$
transmission curves normalized to the peak transmission with the
standard Johnson Cousins functions \citep{Bessell90}.}
\label{figA.1}
\end{figure}

\begin{figure}
\centering
\includegraphics[width=84mm]{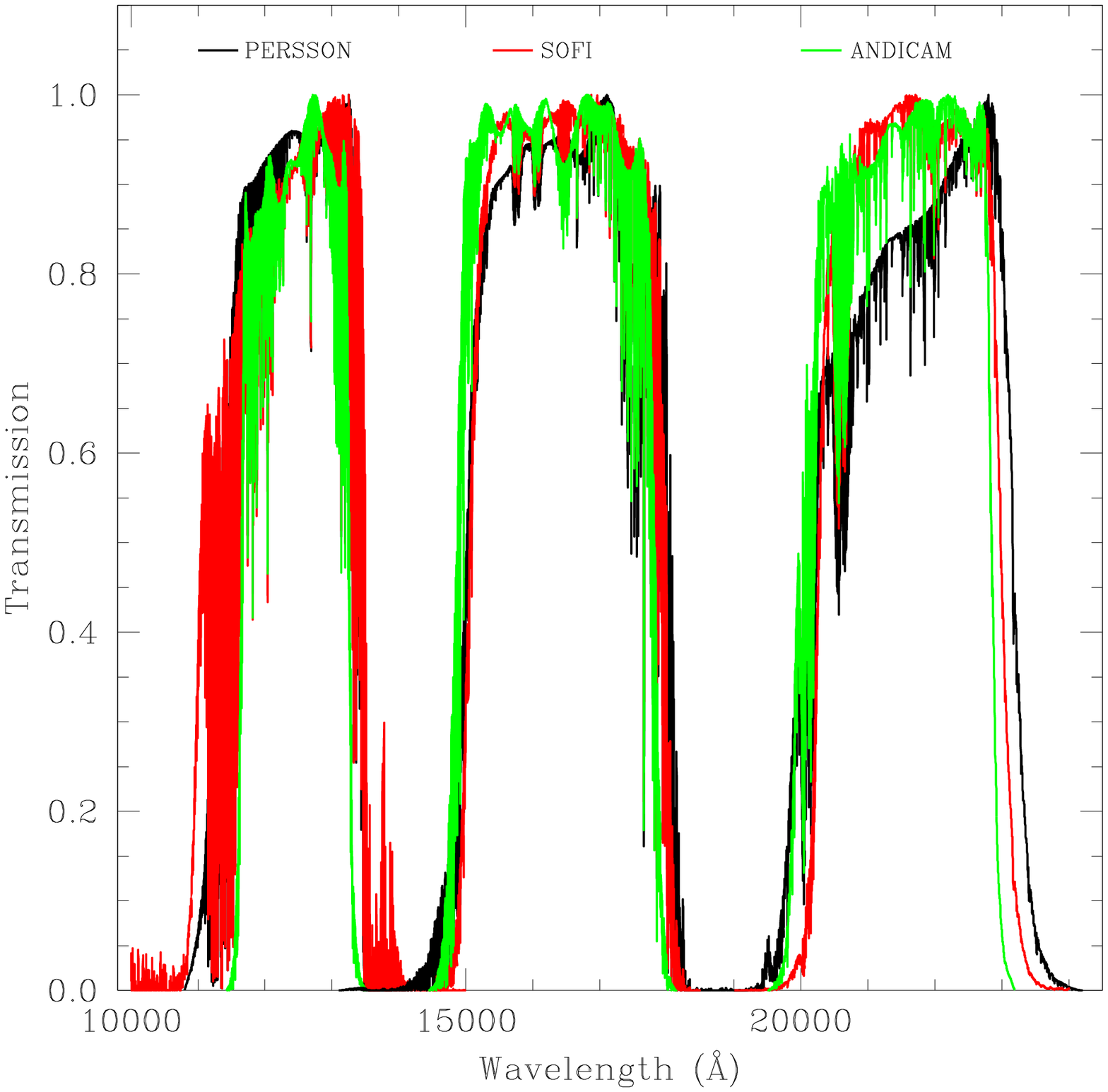}
\caption{Comparison of the different instrumental $JHKs$
transmission curves normalized to the peak transmission with the
standard Persson functions \citep{Persson98}.}
\label{figA.2}
\end{figure}

\begin{table*}
\begin{minipage}{170mm}
\caption{Comparison between synthetic and photometric colour terms.  In
general we used $B$ and $B-V$, $V$ and $B-V$, $R$ and $V-R$ as well as $I$ and
$R-I$. But for EMMI and WFC we use $V$ and $V-R$, for CTIO 0.9m and FORS1 we
use $I$ and $V-I$, and for ALFOSC we use $R$ and $R-I$ and $I$ and $V-I$.  
\label{tabA.1}
}
\tabcolsep 2pt
\begin{tiny}
\begin{tabular}{@{}lllrrrrrrrr}
\hline
 & \multicolumn{1}{c}{sy$^b$} & \multicolumn{1}{c}{ph$^c$} &
  \multicolumn{1}{c}{sy} & \multicolumn{1}{c}{ph} &
  \multicolumn{1}{c}{sy} & \multicolumn{1}{c}{ph} &
  \multicolumn{1}{c}{sy} & \multicolumn{1}{c}{ph} &
  \multicolumn{1}{c}{sy} & \multicolumn{1}{c}{ph} \\
\hline
CTIO 0.9m & 0.062(0.035)$^a$ & 0.098(0.013) & $-$0.092(0.007) &
$-$0.094(0.007) & 0.018(0.001) & 0.016(0.003) & 0.010(0.001) &
$-$0.013(0.009) & 0.004(0.001) & 0.004(0.004)\\
ALFOSC    & 0.093(0.006)$^a$ & 0.076(0.045) & 0.032(0.003) &
0.057(0.016) & $-$0.046(0.002) & $-$0.052(0.007) & $-$0.065(0.002) &
$-$0.072(0.020) &  $-$0.041(0.002) &  $-$0.065(0.029)\\
DFOSC     & 0.058(0.006)$^a$ & 0.029(0.016) & 0.068(0.003) & 
0.089(0.004)  & 0.013(0.002) & 0.013(0.007) & 0.026(0.001) & 
0.009(0.015) & $-$0.048(0.001) & $-$0.060(0.036)\\
JKT       & 0.072(0.050) & 0.026(0.019) & 0.066(0.006) & 0.085(0.032) & 
0.030(0.001) & 0.028(0.035) & 0.003(0.001) & $-$0.039(0.023) & 
0.056(0.004) & 0.018(0.030)\\
WFC       & 0.077(0.021) & 0.059(0.032) & 0.031(0.006) &
\multicolumn{1}{l}{\hspace*{1.9mm}0.063} & $-$0.010(0.001) & 0.007(0.029) &
0.013(0.001) & 0.000(0.014) & $-$0.197(0.004) & $-$0.210(0.004)\\
FORS1     & \multicolumn{1}{c}{$-$} & \multicolumn{1}{c}{$-$} & 
$-$0.083(0.007) & $-$0.083(0.009) & 0.040(0.002) & 0.033(0.010) & 
0.080(0.002) & 0.066(0.010) & $-$0.034(0.004) & $-$0.031(0.005)\\
EMMI      & \multicolumn{1}{c}{$-$} & \multicolumn{1}{c}{$-$} &
0.067(0.006) & $-$0.045(0.005) & 0.017(0.001) & 0.056(0.023) & 
0.067(0.001) & 0.044(0.009) & $-$0.047(0.001) &
\multicolumn{1}{l}{$-$0.038} \\
\hline 
\end{tabular}
\end{tiny}
$^a$ Modified band.\\
$^b$ Colour term computed using synthetic photometry.\\
$^c$ Colour term computed using photometric standard.\\
\end{minipage}
\end{table*}

\begin{table*}
\begin{minipage}{170mm}
\caption{Comparison between synthetic and photometric colour terms for
the infrared instruments. For SofI we use $J$ and $J-K$, $H$ and $J-K$, $Ks$
and $J-K$, while for ANDICAM we use $J$ and $J-Ks$, $H$ and $J-Ks$, $Ks$
and $J-Ks$.
}
\label{tabA.2}
\begin{tiny}
\begin{tabular}{@{}lrrrrrr}
\hline
  & \multicolumn{1}{c}{sy} & \multicolumn{1}{c}{ph} &
  \multicolumn{1}{c}{sy} & \multicolumn{1}{c}{ph} &
  \multicolumn{1}{c}{sy} & \multicolumn{1}{c}{ph}  \\
\hline
SoFi    & $-$0.020 & $-$0.007\hspace*{7.4mm} & $-$0.022 &
0.006\hspace*{7.4mm}  & 0.006 &  0.023\hspace*{7.4mm} \\
ANDICAM    & 0.035 & 0.028(0.005)  & $-$0.009  & $-$0.010(0.005) & 0.000 
& $-$0.003(0.005) \\
\hline 
\end{tabular}
\end{tiny}
\end{minipage}
\end{table*}

\begin{table*}
\begin{minipage}{170mm}
\caption{Optical photometry of SN~2002dj calibrated using the colour equation}
\label{tabA.3}
\begin{tabular}{@{}ccrcccccc}
\hline date & M.J.D. & Phase$^a$ & U & B & V & R & I & Instr. \\
\hline
13/06/2002 & 52439.0 & -11.0  & 16.01 $\pm$  0.03 & 16.02 $\pm$  0.04 & 15.85 $\pm$  0.03 & 15.64 $\pm$  0.06 & 15.96 $\pm$  0.03 & CTIO 0.9m \\
14/06/2002 & 52439.9 & -10.1  &  $-$  & 15.68 $\pm$  0.05 & 15.51 $\pm$  0.02 & 15.25 $\pm$  0.03 & 15.43 $\pm$  0.03 & EMMI \\
27/06/2002 & 52452.9 &   2.9  & 14.30 $\pm$  0.04 & 14.35 $\pm$  0.03 & 14.17 $\pm$  0.02 & 14.09 $\pm$  0.03 & 14.43 $\pm$  0.05 & WFC \\
28/06/2002 & 52453.9 &   3.9  &  $-$  & 14.43 $\pm$  0.03 & 14.17 $\pm$  0.01 & 14.11 $\pm$  0.03 & 14.47 $\pm$  0.03 & JKT \\
29/06/2002 & 52455.0 &   5.0  & 14.40 $\pm$  0.04 & 14.47 $\pm$  0.02 & 14.19 $\pm$  0.02 & 14.14 $\pm$  0.01 & 14.50 $\pm$  0.04 & JKT \\
30/06/2002 & 52456.0 &   6.0  &  $-$  &  $-$  &  $-$  & 14.26 $\pm$  0.03 &  $-$  & JKT \\
02/07/2002 & 52458.0 &   8.0  &  $-$  & 14.68 $\pm$  0.02 & 14.31 $\pm$  0.02 & 14.31 $\pm$  0.03 & 14.64 $\pm$  0.03 & JKT \\
03/07/2002 & 52458.1 &   8.1  & 14.85 $\pm$  0.05 & 14.68 $\pm$  0.05 & 14.32 $\pm$  0.03 & 14.29 $\pm$  0.03 & 14.73 $\pm$  0.05 & DFOSC \\
04/07/2002 & 52460.0 &  10.0  & 15.01 $\pm$  0.02 & 14.83 $\pm$  0.04 & 14.42 $\pm$  0.01 & 14.46 $\pm$  0.04 & 14.96 $\pm$  0.04 & DFOSC \\
07/07/2002 & 52463.0 &  13.0  & 15.42 $\pm$  0.03 & 15.16 $\pm$  0.02 & 14.62 $\pm$  0.03 & 14.61 $\pm$  0.04 & 15.05 $\pm$  0.04 & DFOSC \\
09/07/2002 & 52465.0 &  15.0  & 15.75 $\pm$  0.03 & 15.44 $\pm$  0.04 & 14.79 $\pm$  0.02 & 14.70 $\pm$  0.03 & 15.07 $\pm$  0.04 & DFOSC \\
11/07/2002 & 52467.0 &  17.0  & 16.05 $\pm$  0.01 & 15.61 $\pm$  0.04 & 14.88 $\pm$  0.04 & 14.74 $\pm$  0.05 & 15.04 $\pm$  0.05 & DFOSC \\
16/07/2002 & 52473.0 &  23.0  &  $-$  &  $-$  & 15.14 $\pm$  0.03 & 14.84 $\pm$  0.03 & 14.83 $\pm$  0.02 & CTIO 0.9m \\
16/07/2002 & 52473.1 &  23.1  & 16.75 $\pm$  0.04 & 16.33 $\pm$  0.04 & 15.17 $\pm$  0.03 & 14.79 $\pm$  0.05 & 14.99 $\pm$  0.02 & DFOSC \\
19/07/2002 & 52476.0 &  26.0  & 16.97 $\pm$  0.05 & 16.57 $\pm$  0.04 & 15.42 $\pm$  0.03 & 15.03 $\pm$  0.03 & 14.84 $\pm$  0.04 & DFOSC \\
24/07/2002 & 52479.9 &  29.9  &  $-$  & 16.92 $\pm$  0.05 & 15.54 $\pm$  0.05 & 15.12 $\pm$  0.03 & 14.85 $\pm$  0.01 & ALFOSC \\
25/07/2002 & 52481.0 &  31.0  & 17.41 $\pm$  0.03 & 16.97 $\pm$  0.02 & 15.63 $\pm$  0.03 & 15.20 $\pm$  0.03 & 14.90 $\pm$  0.04 & DFOSC \\
29/07/2002 & 52484.9 &  34.9  & 17.55 $\pm$  0.17 &  $-$  &  $-$  &  $-$  &  $-$  & ALFOSC \\
08/08/2002 & 52495.0 &  45.0  &  $-$  & 17.33 $\pm$  0.04 & 16.22 $\pm$  0.04 & 15.91 $\pm$  0.03 & 15.74 $\pm$  0.04 & CTIO 0.9m \\
31/08/2002 & 52518.0 &  68.0  & 18.38 $\pm$  0.12 &  $-$  & 16.89 $\pm$  0.04 &  $-$  &  $-$  & DFOSC \\
02/09/2002 & 52520.0 &  70.0  &  $-$  &  $-$  & 16.91 $\pm$  0.05 &  $-$  &  $-$  & DFOSC \\
06/09/2002 & 52524.0 &  74.0  &  $-$  & 17.76 $\pm$  0.19 &  $-$  & 17.05 $\pm$  0.16 & 16.88 $\pm$  0.48 & DFOSC \\
25/03/2003 & 52724.3 & 274.3  &  $-$  & 21.00 $\pm$  0.04 & 20.78 $\pm$  0.03 & 21.30 $\pm$  0.08 & 20.74 $\pm$  0.10 & FORS1 \\
\hline
\end{tabular}
 ~$^a$ Relative to the time of the $B$ maximum brightness M.J.D.=52450 $\pm$ 0.7\\
\end{minipage}
\end{table*}

\begin{table*}
\begin{minipage}{110mm}
\caption{Near infrared photometry of SN~2002dj calibrated using the colour equation.}
\label{tabA.4}
\begin{tabular}{@{}ccrcccc}
\hline date & M.J.D. & Phase$^a$ & J & H & K & Instr. \\
\hline
13/06/2002 & 52439.0 & -11.0  & 15.64 $\pm$  0.05 & 15.64 $\pm$  0.08 & 15.66 $\pm$  0.10 & ANDICAM \\
14/06/2002 & 52440.0 & -10.0  & 15.27 $\pm$  0.03 & 15.31 $\pm$  0.05 & 15.25 $\pm$  0.08 & ANDICAM \\
15/06/2002 & 52441.1 &  -8.9  & 14.98 $\pm$  0.03 & 15.13 $\pm$  0.03 & 15.09 $\pm$  0.03 & SoFi \\
19/06/2002 & 52444.1 &  -5.9  & 14.63 $\pm$  0.03 & 14.85 $\pm$  0.03 & 14.70 $\pm$  0.03 & SoFi \\
20/06/2002 & 52446.0 &  -4.0  & 14.56 $\pm$  0.09 & 14.82 $\pm$  0.06 & 14.64 $\pm$  0.06 & ANDICAM \\
24/06/2002 & 52450.0 &  -0.0  & 14.60 $\pm$  0.03 & 14.85 $\pm$  0.05 & 14.50 $\pm$  0.09 & ANDICAM \\
24/06/2002 & 52450.0 &  -0.0  & 14.62 $\pm$  0.03 & 14.83 $\pm$  0.03 & 14.54 $\pm$  0.03 & SoFi \\
27/06/2002 & 52453.1 &   3.1  & 14.88 $\pm$  0.04 & 14.93 $\pm$  0.04 & 14.65 $\pm$  0.07 & ANDICAM \\
30/06/2002 & 52456.0 &   6.0  & 15.33 $\pm$  0.12 & 15.12 $\pm$  0.08 & 14.85 $\pm$  0.16 & ANDICAM \\
08/07/2002 & 52464.0 &  14.0  & 16.38 $\pm$  0.07 & 15.09 $\pm$  0.04 & 15.03 $\pm$  0.08 & ANDICAM \\
11/07/2002 & 52467.0 &  17.0  & 16.36 $\pm$  0.05 & 15.01 $\pm$  0.03 & 14.90 $\pm$  0.05 & ANDICAM \\
11/07/2002 & 52467.0 &  17.0  & 16.15 $\pm$  0.03 & 14.90 $\pm$  0.03 & 14.87 $\pm$  0.03 & SoFi \\
14/07/2002 & 52470.0 &  20.0  & 16.19 $\pm$  0.10 & 14.85 $\pm$  0.04 & 14.76 $\pm$  0.06 & ANDICAM \\
17/07/2002 & 52473.0 &  23.0  & 15.97 $\pm$  0.04 & 14.79 $\pm$  0.03 & 14.66 $\pm$  0.04 & ANDICAM \\
25/07/2002 & 52481.0 &  31.0  & 15.70 $\pm$  0.05 & 14.81 $\pm$  0.04 & 14.83 $\pm$  0.08 & ANDICAM \\
28/07/2002 & 52484.0 &  34.0  & 15.75 $\pm$  0.04 & 15.07 $\pm$  0.04 & 15.13 $\pm$  0.05 & ANDICAM \\
31/07/2002 & 52487.0 &  37.0  & 16.01 $\pm$  0.03 & 15.25 $\pm$  0.06 & 15.45 $\pm$  0.28 & ANDICAM \\
07/08/2002 & 52494.0 &  44.0  & 16.70 $\pm$  0.09 & 15.62 $\pm$  0.07 & 15.67 $\pm$  0.35 & ANDICAM \\
10/08/2002 & 52497.0 &  47.0  & 16.79 $\pm$  0.09 & 15.74 $\pm$  0.05 & 15.94 $\pm$  0.12 & ANDICAM \\
13/08/2002 & 52500.0 &  50.0  & 16.99 $\pm$  0.11 & 15.86 $\pm$  0.05 & 16.05 $\pm$  0.12 & ANDICAM \\
30/08/2002 & 52517.0 &  67.0  & 18.07 $\pm$  0.38 & 16.61 $\pm$  0.03 & 16.87 $\pm$  0.03 & SoFi \\
\hline
\end{tabular}
$^a$ Relative to the time of the $B$ maximum brightness M.J.D.=52450 $\pm$ 0.7\\
\end{minipage}
\end{table*}

\subsection{Band construction} We define the transmission function of a
give passband $S(\lambda)$ as: $S(\lambda)=F(\lambda)\cdot QE(\lambda)
\cdot Ac(\lambda) \cdot Al(\lambda) \cdot M(\lambda) \cdot N_{ref} \cdot
L(\lambda)$\\ where $F(\lambda)$ is the filter transmission function,
$QE(\lambda)$ is the detector quantum efficiency, $Ac(\lambda)$ is the
continuum atmospheric transmission profile, $Al(\lambda)$ is the line
atmospheric transmission profile, $M(\lambda)$ is the mirror
reflectivity function, $N_{ref}$ the number of reflections and
$L(\lambda)$ is the lens throughput.  We downloaded $F(\lambda)$ and
$QE(\lambda)$ from the instrument web sites. For $M(\lambda)$ we used a
standard aluminium reflectivity curve, while we did not find information
about the lens transmission for any of the instruments.  In the
construction of the atmosphere model the weight of the terms
$Ac(\lambda)$ and $ Al(\lambda)$ change considerably going from $U$ to
$Ks$ bands. In the optical the atmosphere transmission is mainly defined
by a continuum function, while in the IR in most of the instruments the
passband cut-offs are defined by deep atmospheric absorption bands. The
latter make the use of a fixed instrumental passband not a good
approximation of the real transmission function, because it depends on
the atmospheric conditions. To a much lesser degree this is also a
problem for the optical $R$ and $I$ filters, where some telluric
features fall in the middle of the passbands. These telluric lines do
not change linearly with airmass, therefore their variation is not
properly accounted by the absorption coefficients.  For La Palma we
obtained $Ac(\lambda)$ from \citet{atm_lapalma}, while for La Silla,
Paranal and CTIO we have used the CTIO transmission curve in the IRAF
reduction package (Stone \& Baldwin 1983, Baldwin \& Stone 1984). For
the IR sky spectrum of La Silla and CTIO we downloaded $Ac(\lambda)
\cdot Al(\lambda)$ from the 2MASS web
side\footnote{\tiny{$http://spider.ipac.caltech.edu/staff/waw/2mass/opt\_cal/south\_atm.tbl.html$}}.
The instrumental filter bands are compared with the standard ones in
Fig.~\ref{figA.1} and Fig.~\ref{figA.2}.\\ As usual (see e.g. Pignata et
al. 2004), we checked how well the constructed instrumental passbands
match the real ones and compared the colour terms derived from standard
stars (\citet{Landolt92} for the optical and \citet{Persson98} for the
IR) with those we computed by integrating the flux of spectrophotometric
standards of \citet{Stritzinger05} for the optical and Sirius, Vega and
the Sun for the IR. We decided to modify the instrumental passband only
in cases where the difference between the two colour terms reported in
Table~\ref{tabA.1} and Table~\ref{tabA.2} were larger than three times
the error associated to both ``synthetic'' and ``photometric'' colour
terms. The lack of a suitable number of spectrophotometric IR standards
with an establish magnitude make the colour term comparison less
effective than at optical wavelengths.

\end{document}